\documentclass[fleqn,usenatbib]{mnras}

\usepackage{newtxtext,newtxmath}

\usepackage[T1]{fontenc}
\usepackage{ae,aecompl}

\usepackage{graphicx}	
\usepackage{amsmath}	

\usepackage{tikz}
\usetikzlibrary{positioning,calc,backgrounds}
\tikzstyle{line}=[draw]
\tikzstyle{arrow}=[draw, -latex] 
\usepackage{siunitx}
\newcommand{\HI}{H\,{\sc i} }
\newcommand{\Mg}{Mg\,{\sc ii} }

\newcommand{\angstrom}{\textup{\AA}}
\newcommand{\note}[1]{\textcolor{black}{#1}}
\newcommand{\report}[1]{\textcolor{black}{#1}}

\title{The distribution and properties of DLAs at $z \leq 2$ in the EAGLE simulations}

\author[Lilian Garratt-Smithson et al.]{
Lilian Garratt -- Smithson,$^{1,2}$\thanks{E-mail: lilian.garratt-smithson@uwa.edu.au} Chris Power,$^{1,2}$ Claudia del P. Lagos,$^{1,2,3}$ 
\newauthor Adam R.~H. Stevens,$^{1,2}$ James R. Allison,$^{2,4}$ and Elaine M. Sadler.$^{2,5,6}$\\
$^{1}$International Centre for Radio Astronomy Research,University of Western Australia, 35 Stirling Highway, Crawley, Western Australia 6009, Australia\\
$^{2}$ARC Centre of Excellence for All-sky Astrophysics in 3 Dimensions (ASTRO 3D)\\
$^{3}$Cosmic Dawn Center (DAWN)\\
$^{4}$Sub-Department of Astrophysics, Department of Physics, University of Oxford, Denys Wilkinson Building, Keble Rd., Oxford OX1 3RH, UK\\
$^{5}$Sydney Institute for Astronomy, School of Physics A28, University of Sydney, Sydney, NSW 2006, Australia\\
$^{6}$CSIRO Astronomy and Space Science, PO Box 76, Epping, NSW 1710, Australia\\
}

\date{Accepted XXX. Received YYY; in original form ZZZ}

\pubyear{2020}

\begin{document}
\label{firstpage}
\pagerange{\pageref{firstpage}--\pageref{lastpage}}
\maketitle

\begin{abstract}
Determining the spatial distribution and intrinsic physical properties of neutral hydrogen on cosmological scales is one of the key goals of next-generation radio surveys. We use the EAGLE galaxy formation simulations to assess the properties of damped Lyman-alpha absorbers (DLAs) that are associated with galaxies and their underlying dark matter haloes between $0 \leq z \leq 2$. We find that the covering fraction of DLAs increases at higher redshift; a significant fraction of neutral atomic hydrogen (H\,{\sc i}) resides in the outskirts of galaxies with stellar mass \note{$\geq$ 10$^{10}$ M$_{_{\rm{\odot}}}$}; and the covering fraction of DLAs in the \note{circumgalactic medium (CGM)} is enhanced relative to that of the \note{interstellar medium (ISM) with increasing halo mass}. Moreover, we find that the mean density of the \HI in galaxies increases with increasing stellar mass, while the DLAs in high- and low-halo-mass systems have higher column densities than those in galaxies with intermediate halo masses ($\sim$ 10$^{12}$ M$_{\odot}$ at $z$~$= 0$). These high-impact CGM DLAs in high-stellar-mass systems tend to be metal-poor, likely tracing smooth accretion. Overall, our results point to the CGM playing an important role in DLA studies at high redshift ($z$~$\geq 1$). However, their properties are impacted both by numerical resolution and the detailed feedback prescriptions employed in cosmological simulations, particularly that of AGN.

\end{abstract}



\section{Introduction}
Hydrogen is the most abundant element in the \note{Universe} and \HI (atomic hydrogen) has been linked with fundamental galaxy properties such as \note{star formation rate (SFR)} and \note{star formation efficiency (SFE)}, stellar mass, morphology, metallicity and colour \citep[e.g.][]{Zhang2009, Cortese2011, Wang2011, Hughes2013, Saintonge2016, Wang2017, Zhou2018}. Additionally, the column density of \HI is also known to increase towards the centre of galaxies \citep[e.g.][]{Rahmati2014, Prochaska2017, Rhodin2018}. \note{These correlations imply} an intrinsic link between \HI and internal galactic processes. 

There existed at least twice as much \HI in the high-redshift Universe ($z$~$\sim$ 2) than today \citep[e.g.][]{Prochaska2009, Neeleman2016, Sanchez-Ramirez2016, Bird2017, Rhee2018}. However, the evolution of the cosmic \HI mass density ($\Omega_{\rm{HI}}$) is relatively weak compared with the evolution of $\Omega_{\rm{H_2}}$ (the cosmic molecular hydrogen mass density), which declines by a factor of 3--10 from $z$~$\sim 2$ to 0 \citep{Decarli2019}, along with $\Omega_{\rm{SFR}}$ (the cosmic SFR density), which shows a sharp peak at $z$~$\sim$ 2 \citep{Madau2014, Driver2018} and drops by an order of magnitude ($\sim$ 20$\times$) to the present-day Universe. This suggests \HI is replenished throughout cosmic time but that the conversion efficiency into H$_2$ and SFR is evolving. Recent studies have indicated that the \HI reservoir inside a galaxy to be highly dynamic and existing in a state of flux, accreting onto a galaxy initially as ionized inflows and fuelling further star formation on Gyr timescales \citep[e.g.][]{Dave2013, Crain2017}. \report{\citet{Chowdhury2020} recently used \HI 21-cm emission stacking to obtain an $\Omega_{\rm{HI}}$ value that is consistent with previous optical measurements using \Mg absorbers and UV DLAs, while also finding the average M$_{\rm{HI}}$-to-SFR ratio suggests a relatively short \HI depletion timescale (of the order of a few Gyr).}

\HI ubiquitously emits and absorbs at a wavelength of 21-cm. 21-cm absorption line studies are an important tool for studying \HI in the early Universe, particularly since the sensitivity of absorption line surveys is independent of redshift. This is in contrast to \HI emission, which is \note{very weak for distant galaxies \citep[e.g.][]{Fernandez2016}}. Practically, to observe 21-cm absorption requires a background source that is bright in the radio. This is normally a quasar. 

The next generation of blind 21-cm absorption surveys will open up a new area of parameter space \report{\citep[for a recent example, see work by][who use \HI 21 cm absorption in the redshift range $0<z\leq2.74$ to place constraints on the cosmic evolution of $\Omega_{\rm{HI}}$]{Grasha2020}.} In particular, both the MeerKAT Absorption Line Survey \citep[MALS;][]{Gupta2016} and the First Large Absorption Survey in \HI \citep[FLASH;][]{Allison2020} are two dedicated 21-cm absorption surveys at cosmological distances. Here we focus on FLASH, which is a blind 21-cm survey of the entire southern sky, probing \note{150,000} sightlines in the redshift range $0 - 1$ with the Australian Square Kilometre Array Pathfinder \citep[ASKAP;][]{Johnston2007}. 

FLASH will be sensitive to the highest column density \HI gas \citep[see Fig. 6 of][]{Allison2016} and hence most likely DLAs \citep[damped Lyman-alpha systems, defined as systems with N$_{\rm{HI}}$ values of greater than 2 $\times$ 10$^{20}$ cm$^{-2}$; for a review see][]{Wolfe2005}. DLAs are thought to typically arise in the inner 30 kpc of a galaxy, since column density appears to anti-correlate with impact parameter \citep[e.g.][]{Rahmati2014, Rubin2015}. In this way, DLAs \note{are expected to} trace the crucial star formation processes inside a galaxy, in particular the cooling of atomic to molecular hydrogen, which is the fuel for further star formation. Additionally, 21-cm absorption is sensitive to the harmonic mean 21-cm excitation temperature, which in turn is dependent on the fraction of the CNM \citep[cold neutral medium; e.g.][]{Kanekar2014, Allison2016}. Therefore, 21-cm surveys are sensitive to the coldest gas in the galaxy, unlike optical DLA searches that trace the total \HI content.

Recently, \citet{Allison2020} demonstrated early science results from FLASH by carrying out a blind survey for 21-cm absorption in the GAMA 23 field \citep{Liske2015}. They detected absorption in the outskirts (impact parameter $= 17\,$kpc) of an intervening early type galaxy, showing that substantial column densities of cold absorbing gas can be found at large distances (presumably in an \HI disc) in such galaxies. Additionally, in ASKAP commissioning observations of a sample of radio-loud quasars, \citet{Sadler2020} obtained a measurement of the incidence of DLAs at $0.4 < z < 1$ that was consistent with previous measurements within the current uncertainties. These measurements will improve by two orders of magnitude once FLASH is complete

\note{Due to observational limits (i.e. the use of optical telescopes) DLAs have been well studied at $z$~$> 2$, the range where the rest-frame UV Lyman-alpha line is redshifted into the optical. Surveys such as the Sloan Digital Sky Survey \citep[SDSS;][]{York2000, Noterdaeme2009, Noterdaeme2012} have been critical to provide a range of valuable observations. However, at $z$~$< 1$, it has been particularly difficult to obtain a sample of unbiased DLAs. The common approach has been to use SDSS to pre-select \Mg absorbers \citep[e.g.][]{Turnshek2005} and then follow these up with space UV spectrographs, such as the Hubble Space Telescope \citep[for a recent example see][]{Monier2019}. It is, however, unclear whether this approach delivers a representative sample of intermediate-redshift DLAs, with some tension in the derived $\Omega_{\rm HI}$ seen in the literature \citep[e.g.][]{Neeleman2016, Berg2017, Rao2017}. FLASH will observe DLAs in the radio towards radio-bright galaxies, and hence will provide the first unbiased view of neutral hydrogen in the Universe at intermediate redshifts.}

\note{From observational studies, we still have a limited understanding of DLA host galaxies, primarily due to the lack of follow-up detections in multi-wavelength emission \citep[e.g.][]{Fynbo2011, Peroux2011, Fumagalli2015, Rahmani2016}. The lack of follow-up detections is attributed both to contamination from the background light of the quasar \citep[e.g.][]{Moller1998}, along with the fact DLAs are thought to trace galaxies with a wide range of galaxy properties, which makes them a more representative sample of the overall galaxy population \citep[e.g.][]{Krogager2017}. It also means a large number of DLAs are likely to be associated with faint, low mass galaxies that are below the emission detection limit \citep[e.g.][]{Fumagalli2015}. There have been a number of observational papers that look at the properties of DLAs, in particular their metallicities \citep[e.g.][]{Rafelski2014, Krogager2017} and kinematics \citep[e.g.][]{Prochaska1997}, but their physical nature is still debated. }

Hydrodynamical simulations, alongside semi-analytic models of galaxy formation, offer a way to theoretically investigate the physical origin of DLAs and their host galaxy properties, free from observational selection effects \citep[examples of simulations/models investigating DLAs include][]{Altay2013, Berry2014, Rahmati2015, Berry2016, Rhodin2019, VandeVoort2019}. There has been a large number of previous works investigating the number density of DLAs \note{in cosmological hydrodynamical simulations} and how this varies with properties of the host galaxy such as stellar mass and star formation rate, along with the range of impact parameters possible \citep[e.g.][]{Bird2014, Rahmati2014}. Similar to the observations, this work has typically been focused at $z$~$\geq 2$, and simulations/models suggest DLAs trace a representative sample of galaxies. Moreover, the majority are considered to be hosted by faint low-halo-mass systems \note{and thus have low} star formation rates \citep[e.g.][]{Bird2014, Rahmati2014}. Work using theoretical models has also suggested that the physical origin of DLAs is redshift-dependent, where DLAs of $z$~$> 3$ are likely to arise from intergalactic gas filaments, while those below this redshift are likely to originate from the galactic disc \citep[][]{Berry2014}. 

Due to the high computational demand of resolving the cold neutral ISM (interstellar medium) of galaxies, a number of recent theoretical studies have used cosmological zoom simulations \report{\citep[e.g.][]{Rhodin2019}}, which can resolve cooling below 10$^4$ K and hence model \HI self-consistently throughout the simulation. In particular, simulations such as these have shed light on the contribution of the CGM (circumgalactic medium)/ gaseous halo to the DLA population. However these smaller scale simulations lose the large number statistics required to compare \note{with surveys}.

In order to interpret the results from FLASH, it is important to make predictions about (a) the distribution of \HI in and around galaxies at this redshift range as a function of their properties (e.g. halo mass, stellar kinematics) and (b) the properties of DLAs associated with galaxies in this redshift range. In this paper we hope to address both questions by utilising the Evolution and Assembly of GaLaxies and
their Environments (EAGLE) cosmological simulations, \citep{Schaye2015}.

Previous work has shown EAGLE boxes of varying resolution are able to largely reproduce a wide range of cold gas properties (despite not been calibrated to do so), including the observed clustering of \HI systems \citep[][]{Crain2017}, the global \HI column density distribution function \citep{Rahmati2015}, the observed \HI morphologies of galaxies and in particular their radial profiles \citep{Bahe2016}, along with their H$_{\rm{2}}$ properties \citep{Lagos2015, Lagos2016}. For a recent detailed analysis of the cold gas contents in EAGLE and other cosmological simulations, see \citet{Dave2020}.

In this paper we are particularly interested in investigating further the distribution of strong DLAs in galaxies with a wide range of halo masses in the redshift range of interest for FLASH. We are also interested in the relative contribution of the CGM and how this varies with galaxy properties. Moreover, we will discuss the DLA properties (such as metallicity and impact parameter) and how these vary with host galaxy properties and redshift. 

The structure of the paper will be as follows; the next section will explain our numerical methodology, in particular detailing how we obtain the \HI mass fractions from EAGLE, along with our method of separating gas particles into the ISM/CGM (Section \ref{sec:num_method}). Our results will follow in Section \ref{sec:results}, beginning with the DLA covering fractions in EAGLE galaxies and moving on to the detailed DLA properties. Finally we give a discussion and our conclusions in Section \ref{sec:discuss_and_concl}.
\section{Numerical Method}\label{sec:num_method}
\subsection{The EAGLE simulation\note{s}}
In this paper, we use the public data release of the EAGLE simulation suite \citep{Crain2015, Schaye2015, McAlpine2016, Eagle2017}. EAGLE is a set of cosmological hydrodynamical simulations run using a modified version of the N-body/Smoothed Particle Hydrodynamics (SPH) code GADGET 3 \citep{Springel2005}. The modifications to the SPH method are collectively known as `Anarchy' and include an artificial viscosity switch \citep{Cullen2010}, an artificial thermal conduction switch \citep{Price2008}, a time-step limiter \citep{Durier2012} and the pressure-entropy formulation of \citet{Hopkins2013}. 

Each simulation also includes a wide range of sub-grid physics, including tracking stellar winds from Asymptotic Giant Branch (AGB) stars and supernovae \citep[core-collapse and Type 1a,][]{Wiersma2009}, along with a stochastic star formation recipe \citep{Schaye2008} that also includes a metal-dependant star formation threshold based on \note{\citet{Schaye2004}} and is designed to take into account the atomic-molecular transition. Moreover, radiative cooling and heating are included \citep{Wiersma2009b}, along with stellar feedback via a stochastic thermal heating \citep[see ][for details]{DallaVecchia2012} and an \note{active galactic nuclei (AGN)} feedback scheme based on a modified Bondi-Hoyle accretion rate \citep{RosasGuevara2015} and using the energy threshold described in \citet{Schaye2015}. The subgrid parameters were calibrated to reproduce the $z$~$= 0$ galaxy stellar mass function (GSMF), along with the galaxy stellar mass--size relation and the black hole--stellar mass relations. However, the gaseous properties of the galaxies were not calibrated and hence \note{relations that include the gas properties of galaxies can be considered as predictions}.

For \report{the majority of results in this paper,} we use simulations run at two different volumes; (100 cMpc)$^3$ (Ref-L0100N1504, referred to as RefL0100 in this paper) and (25 cMpc)$^3$ (Recal-L0025N0752, referred to as RecalL0025). Our choice was motivated by the intrinsic link between the cold gas fractions of galaxies and the detailed numerical modelling of physical processes such as stellar feedback, which are implemented differently in the two boxes (see below for details). We also use both boxes in order to quantify the effects of numerical resolution on our results. The smaller EAGLE box also allows us to include a greater number of lower-stellar-mass galaxies ($< 10^{10}$ M$_{\odot}$), given these are adequately resolved in the RecalL0025 box. This is advantageous since previous studies \citep[e.g.][]{Bird2014, Rahmati2014} predict the majority of DLAs to be in low mass haloes (M$_{\rm{200}}$ $<$ 10$^{10}$ M$_{\rm{\odot}}$), while there is also evidence for strong DLAs in galaxies of halo mass 10$^{11}$ M$_{\rm{\odot}}$ -- 10$^{12}$ M$_{\rm{\odot}}$ at $z$~$> 3$ \citep{Mackenzie2019}. Furthermore, \citet{Crain2017} found the standard-resolution box (Ref L0100) was unable to reproduce the \HI CDDF (column density distribution function) \note{due to systematically underestimating \HI column densities \citep[see Fig. 2 of][]{Crain2017}}. The authors also found this was largely corrected for in the higher resolution box.

The two models here utilise different values for four key sub-grid parameters \citep[described in Table 3 of ][]{Schaye2015}, relating to both the stellar feedback, AGN feedback and the black hole (BH) accretion rate. These were altered for the higher-resolution run in order to obtain better agreement with the $z$~$= 0$ GSMF, thereby achieving `weak convergence' \citep[for a discussion on this see section 2.2 of ][]{Schaye2015} with the standard-resolution run. 

\report{During this paper, we also refer to two EAGLE simulations run using a box size of 50 cMpc$^3$, with/without AGN feedback turned on; RefL0050N0752 and NoAGNL0050N0752 respectively. These simulations are used to isolate the impact of AGN feedback on our results.}
\subsection{Our sample of galaxies}
This work uses the halo catalogue provided by the EAGLE collaboration \citep{McAlpine2016} in order to identify haloes of interest. These haloes were identified using the Friends-of-Friends (FoF) method \citep{Davis1985} and SUBFIND algorithms \citep{Springel2001, Dolag2009}. As in \citet{Bahe2016}, we chose to focus our study on central galaxies in order to avoid disentangling the array of environmental processes undergone by satellite galaxies. Such a study is beyond the scope of this paper, however \note{\citet{Marasco2016} investigated the link between \HI and environment for satellite galaxies in EAGLE.} 

We use the values \note{for R$_{\rm{200c}}$ (defined as the radius within which density is 200 times the critical density of the Universe) and M$_{\rm{200c}}$ (the mass within R$_{\rm{200c}}$) that were computed using the SUBFIND algorithms.} The centre of the halo is taken as the coordinate of the particle with the minimum potential and M$_{\rm{200c}}$ (also referred to as M$_{200}$ within the paper) is the halo virial mass. We also apply a stellar mass cut of 10$^{10}$ M$_{\rm{\odot}}$ to the galaxies in RefL0100 and a cut of \note{10$^9$ M$_{\rm{\odot}}$} to the galaxies in RecalL00025. This ensures the galaxies we study are adequately resolved, particularly when exploring the stellar kinematic morphology of individual galaxies. Throughout the paper all distances quoted in kpc are proper distances (i.e. pkpc), not comoving.

\subsection{Calculating galaxy properties}
\subsubsection{\note{Atomic/molecular hydrogen breakdown}}\label{sec:HI_method}
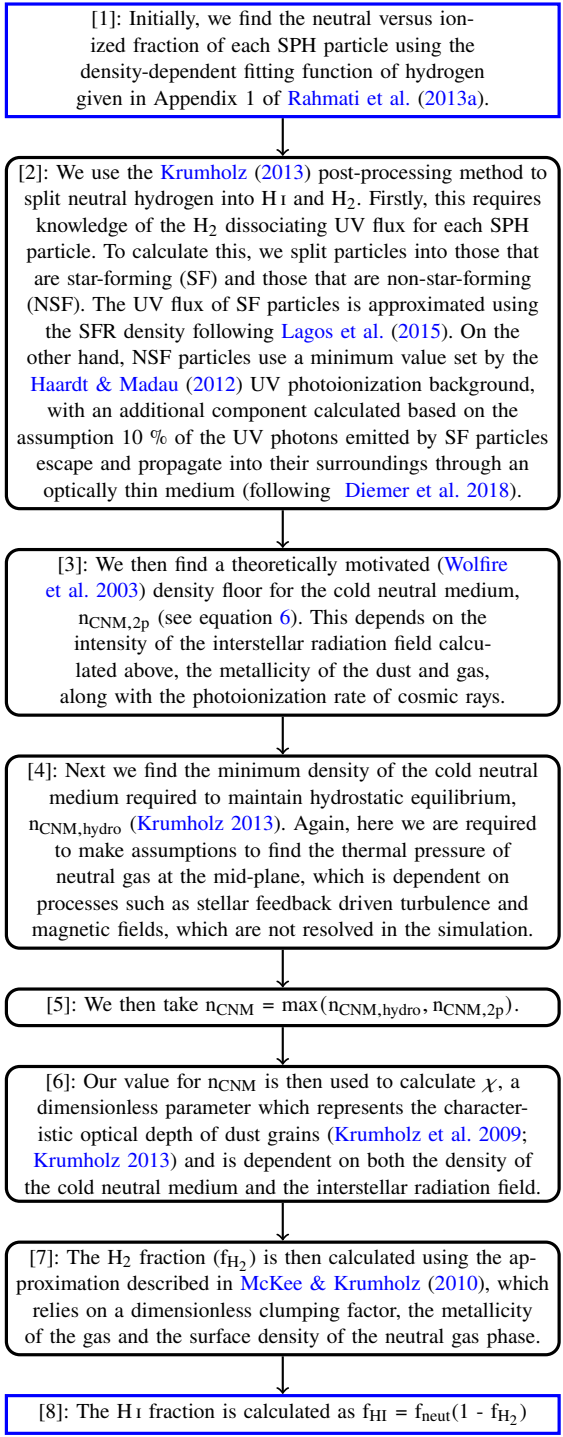
\begin{figure}\label{fig:flow_diagram}
\centering
\begin{tikzpicture}[
    roundnode/.style={rectangle, rounded corners = 5pt, draw=black, very thick, text width=0.4\textwidth,align=center},
    squarednode/.style={rectangle, draw=blue, very thick, minimum size=5mm, text width=0.4\textwidth, align=center},
    node distance=5mm
    ]
    \node[squarednode](init){
   [1]: Initially, we find the neutral versus ionized fraction of each SPH particle using the density-dependent fitting function of hydrogen given in Appendix 1 of \citet{Rahmati2013a}. 
    };
    \node[roundnode](uv)[below=of init]{
    [2]: We use the \citet{Krumholz2013} post-processing method to split neutral hydrogen into \HI and H$_{2}$. Firstly, this requires knowledge of the H$_{2}$ dissociating UV flux for each SPH particle. To calculate this, we split particles into those that are star-forming (SF) and those that are non-star-forming (NSF). The UV flux of SF particles is approximated using the SFR density following \citet{Lagos2015}. On the other hand, NSF particles use a minimum value set by the \citet{Haardt2012} UV photoionization background, with an additional component calculated based on the assumption 10 $\%$ of the UV photons emitted by SF particles escape and propagate into their surroundings through an optically thin medium \citep[following ][]{Diemer2018}.
    };
    \node[roundnode](ncnm2p)[below=of uv]{
    [3]: We then find a theoretically motivated \citep{Wolfire2003} density floor for the cold neutral medium, n$_{\rm{CNM,2p}}$ (see equation \ref{eqn:nCNM_2p_simplified}). This depends on the intensity of the interstellar radiation field calculated above, the metallicity of the dust and gas, along with the photoionization rate of cosmic rays. 
    };
    \node[roundnode](ncnmhydro)[below=of ncnm2p]{
    [4]: Next we find the minimum density of the cold neutral medium required to maintain hydrostatic equilibrium, n$_{\rm{CNM, hydro}}$ \citep{Krumholz2013}. Again, here we are required to make assumptions to find the thermal pressure of neutral gas at the mid-plane, which is dependent on processes such as stellar feedback driven turbulence and magnetic fields, which are not resolved in the simulation.
    };
    \node[roundnode](ncnm)[below=of ncnmhydro]{
    [5]: We then take n$_{\rm{CNM}}$ = $\rm{max(n_{CNM, hydro}, n_{CNM, 2p})}$. 
    };
    \node[roundnode](chi)[below=of ncnm]{
    [6]: Our value for n$_{\rm{CNM}}$ is then used to calculate $\chi$, a dimensionless parameter which represents the characteristic optical depth of dust grains \citep{Krumholz2009, Krumholz2013} and is dependent on both the density of the cold neutral medium and the interstellar radiation field.
    };
    \node[roundnode](H2)[below=of chi]{
    [7]: The H$_{\rm{2}}$ fraction (f$_{\rm{H_2}}$) is then calculated using the approximation described in \citet{McKee2010}, which relies on a dimensionless clumping factor, the metallicity of the gas and the surface density of the neutral gas phase.
    };
    \node[squarednode](HI)[below=of H2]{
    [8]: The \HI fraction is calculated as f$_{\rm{HI}}$ = f$_{\rm{neut}}$(1 - f$_{\rm{H_2}}$)
    };

    \draw[thick,->] (init.south) -- (uv.north);
    \draw[->,thick] (uv.south) -- (ncnm2p.north);
    \draw[->,thick] (ncnm2p.south) -- (ncnmhydro.north);
    \draw[->,thick] (ncnmhydro.south) -- (ncnm.north);
    \draw[->,thick] (ncnm.south) -- (chi.north);
    \draw[->,thick] (chi.south) -- (H2.north);
    \draw[->,thick] (H2.south) -- (HI.north);
\end{tikzpicture}
\caption{The process of \HI/H$_2$ breakdown, as described in Section \ref{sec:HI_method}.}
\end{figure}

While EAGLE models hydrogen, helium and nine metals self-consistently, the mass resolution of the simulations is insufficient to follow cold gas to form atoms and molecules (a temperature floor of 8000 K is imposed to avoid artificial fragmentation). Therefore the ionized/neutral fractions of hydrogen, along with the fraction of molecular hydrogen (H$_{2}$) needs to be calculated in post-processing. \note{Our method, detailed in this section, is summarised in Fig. \ref{fig:flow_diagram}}. Beyond this, those who are less interested in our detailed \HI -- H$_{2}$ breakdown method can skip to Section \ref{sec:HI_kinematics_method}, where we go on to talk about our \HI kinematic galaxy classifications.

\label{fig:ORBflow}
Our method follows previous works \citep[e.g][]{Crain2017, Bahe2016, Lagos2016, Lagos2015}, calculating the \HI mass on a particle-by-particle basis, \note{however we also include modifications based on the method outlined in \citet{Diemer2018}}. Firstly, the neutral/ionized fraction of hydrogen of each SPH particle was found. To do so, we used the density-dependent fitting function for the photoionization rate of hydrogen given in Appendix~A1 of \cite{Rahmati2013a}. This was formulated based on cosmological simulations between $z$~$= 0 - 5$ which included radiative transfer (RT) calculations. We linearly interpolate non-integer redshifts, finding our results are insensitive to the linear interpolation method. In their \citeyear{Rahmati2013a} paper, \citeauthor{Rahmati2013a} find good agreement between the photoionization rates obtained via post-processing simulations using this best-fitting function and the results of the full RT calculations.

Furthermore, the \cite{Rahmati2013a} fitting formula gives the total photoionization rate, $\Gamma_{\rm{pho}}$, as a fraction of the ionization rate due to background ionizing radiation. We therefore calculated the redshift, temperature and density-dependent value for the background photoionization rate ($\Gamma_{\rm{UVB}}$) based on the publicly available tables given by \cite{Haardt2012}\footnote[1]{available at: \url{http://www.ucolick.org/~pmadau/CUBA/HOME.html}}. These models were calculated using the RT code CUBA \citep{Haardt1996,Haardt2001} and represent an updated version of previous models \citep{Madau1995,Haardt1996,Madau1999}. We then used this value of $\Gamma_{\rm{UVB}}$, along with best fit parameters linearly interpolated using Table A1 of \cite{Rahmati2013a}, to calculate $\Gamma_{\rm{pho}}$. This was then fed into the method outlined in Appendix~A2 of \cite{Rahmati2013a} in order to arrive at the fraction of neutral hydrogen for each gas particle. As in previous works \citep[e.g.][]{Lagos2015, Crain2017} we set the temperature of star-forming particles to 10$^4$ K, since the temperature of these particles is not physical and instead set by an imposed polytropic equation of state (where $P_{\rm{eos}} \propto \rho_{\rm{g}}^{4/3}$), which limits artificial fragmentation. 10$^4$ K was chosen to mimic the warm, diffuse ISM surrounding young stellar populations \citep{Crain2017}.

It should be noted that the EAGLE simulations were run using the older photionization rate tables presented in \citet{Haardt2001}. However, we expect this not to affect our results since the gas we are interested in has a column density above that required for self-shielding (N$_{\rm{HI}}$). Moreover, \citet{Rahmati2015} showed that using the \citet{Haardt2012} model, with its associated reduction in the photoionization rate of hydrogen, improved the agreement between the abundance of \HI absorbers in EAGLE and the abundance observed at high redshift (z $= 2.5$).
 
\citet{Rahmati2013b}, showed that local sources of photo-ionizing feedback have a significant impact on the dense (N$_{\rm{HI}} \sim$ 10$^{18}$ -- 10$^{21}$ cm$^{-2}$) \HI gas distribution in the galaxy. This effect was less significant below N$_{\rm{HI}} \sim$ 10$^{20.5}$ cm$^{-2}$ at the redshift range we are interested in \citep[z $< 1$; see Fig. 9 of ][]{Rahmati2013b}, however could significantly impact the CDDF of strong DLAs. Furthermore, \citet{Rahmati2013b} found the impact of \note{local stellar radiation (LSR)} on strong DLA systems is highly uncertain due to its dependence on the complex morphology of the ISM on local scales. Therefore, although the coupling of the photo-ionizing radiation from the LSR to DLAs is relevant to this work, it is beyond the scope if this paper. We do, however, take into account the molecular-dissociating radiation associated with LSR and this will be detailed below. Since the photo-ionizing radiation is not included, we take N$_{\rm{HI}} >$ 10$^{21}$ cm$^{-2}$ as an upper limit. We discuss the prevalence of strong DLAs in our simulations in sections \ref{sec:fcov_vs_fvol} and \ref{sec:discussion}.

There are a number of different post-processing methods previously employed to split neutral hydrogen into \HI and H$_2$ in cosmological simulations. \note{A} recent review of the subject is included in \citet{Diemer2018}. For the majority of the results in this paper we chose to follow the method outlined in \cite{Krumholz2013}, as has been previously utilised for cosmological simulations \citep[e.g.][]{Lagos2015, Diemer2018, Stevens2019}. This prescription divides the neutral atomic \HI gas of a galactic disc into a cold (T < 300 K) and \note{warm} (T $\sim$ 10$^{4}$ K) phase, alongside a gravitationally bound H$_{\rm{2}}$/ molecular phase. It is important to note here that a thermally unstable medium, or UNM, also exists, with spin temperatures of 250 K to 1000 K and constituting 20$\%$ of the total \HI mass \citep{Murray2018}. However, the inclusion of this phase is beyond the scope of the paper.

\cite{Wolfire2003} showed the cold and hot \HI phase can exist in thermal equilibrium across a narrow range of pressures and that the cold neutral ISM (CNM) can have a maximum temperature, $T_{\rm{CNM,max}}$$\approx$ 243 K, which imposes a corresponding density floor, n$_{\rm{CNM,2p}}$, described by
\begin{equation}\label{eqn:nCNM_2p}
    n_{\rm{CNM,2p}} \approx 31 G_0' \frac{Z_{\rm{d}}/ Z_{\rm{g}}}{1 + 3.1(G_0' Z_{\rm{d}}/\xi_t')},
\end{equation}
where $G_0'$ is the intensity of the interstellar radiation field (ISRF) in units of the Habing radiation field (defined in the wavelength range $912-2400 \angstrom$), $Z_d$ and $Z_g$ are the metallicities of the dust and gas respectively and $\xi_t'$ is the ionization rate due to cosmic rays and X-rays \citep{Krumholz2013,Wolfire2003}. Here for star-forming gas particles we follow \cite{Lagos2015}, where $G_0'$ is approximated as the ratio between the SFR surface density of the gas and the SFR surface density in the solar neighborhood, which we also take to be $10^{-3}$ M$_{\rm{\odot}}$ yr$^{-1}$ kpc$^{-2}$. To find $\Sigma_{\rm{SFR}}$ we calculated the SFR density, $\rho_{\rm{SFR}}$, using the instantaneous SFR outputted by EAGLE, $\dot{m}_{\rm{\star}} $\citep[equation 1 from ][]{Schaye2015}, along with the relation $\dot{m}_{\rm{\star}} = m_g (\rho_{\rm{SFR}} / \rho_g)$. We then multiply $\rho_{\rm{SFR}}$ by the Jeans length, $\lambda_{\rm{J}}$, to obtain $\Sigma_{\rm{SFR}}$. Here $\lambda_{\rm{J}}$ is defined by
\begin{equation}\label{Jeans}
    \lambda_J = \frac{c_{\rm{s,eff}}}{\sqrt[]{G \rho_{\rm{H}}}},
\end{equation}
\citep{Lagos2015}. \note{Where} $\rho_{\rm{H}}$ is the hydrogen density and $c_{\rm{s,eff}}$ is the effective sound speed, given by
\begin{equation}
    c_{\rm{s,eff}} = \sqrt[]{\frac{\gamma P_{\rm{tot}}}{\rho_{\rm{g}}}},
\end{equation}
\citep{Schaye2008} where $P_{\rm{tot}}$ is the total mid plane pressure and $\gamma$ is the ratio of specific heats. In order to calculate $\gamma$ (and the mean molecular weight $\mu$), both of which are dependent on the atomic-to-molecular ratio (f$_{\rm{mol}}$), we follow the iterative method outlined in \citet{Stevens2019}, where
\begin{equation}
   \gamma = \frac{5}{3} (1 - f_{\rm{mol}}) + \frac{7}{5} f_{\rm{mol}},
\end{equation}
and
\begin{equation}
    f_{\rm{mol}} =\frac{X f_{\rm{neut}} \, f_{\rm{H_2}}}{1 - Z},
\end{equation}
where $X$ is the total hydrogen mass fraction, $f_{\rm{neut}}$ is the neutral mass fraction calculated above and $f_{\rm{H_2}}$ is the fraction of the neutral hydrogen which is molecular \citep{Stevens2019}.

On the other hand, the minimum value of $G_0'$ for each non-star-forming particle is set to the UV photoionization background of \citet{Haardt2012}. On top of this, we follow \citet{Diemer2018} by assuming 10$\%$ of the UV photons emitted by star-forming particles escape and propagate through their surroundings via an optically thin medium. This method of modelling the UV strength in non-star-forming particles, detailed in \citet{Diemer2018}, avoids a sharp transition between the star-forming ISM and the non-star-forming ISM. As in \citet{Diemer2018}, we used a fast Fourier transform (FFT) technique to convolve the SFR distribution and a 1/r$^2$ Green's function using two 128$^3$ grids -- one taking into account the entire galaxy, the second centered on 2 gas half-mass radii -- and then interpolated the UV flux at the position of each gas particle. For non-star-forming gas particles inside a given grid cell, a $1 / \sqrt{3 r_{\rm{cell} / 2}^3}$ contribution from the SFR particles is assumed \citep{Diemer2018}.

Equation~\ref{eqn:nCNM_2p} is simplified in \cite{Krumholz2013} by making the approximation $Z_{\rm d} = Z_{\rm g}$, based on the fact both are attributed to a common supply of metals, along with the substitution $G_0'/\xi_t' = 1$, which relies on the fact both scale with SFR. Performing these substitutions yields
\begin{equation}\label{eqn:nCNM_2p_simplified}
    n_{\rm{CNM,2p}} \approx 23 G_0' \left(\frac{1 + 3.1(Z/Z_{\rm{\odot}})^{0.365}}{4.1}\right)^{-1},
\end{equation}
\citep{Krumholz2013}. However, equation \ref{eqn:nCNM_2p_simplified} implies in the absence of an stellar radiation field $n_{\rm{min, CNM}} \rightarrow 0$, along with the pressure of the CNM. This unphysical consequence is avoided if one considers the minimum CNM density to maintain hydrostatic equilibrium. Here, the pressure of the neutral gas in the galactic disc ($P_{\rm{th, disc}}$) can be split into three terms describing the self-gravity of the \HI which is not bound (1st term), along with the gravitational interaction between the \HI and the molecular clouds (2nd term) and the surrounding stars and dark matter (3rd term); 
\begin{equation}\label{eqn:disc_pressure}
    P_{\rm{th,disc}} = \frac{\pi}{2}G\Sigma_{\rm{HI}}^2 + \pi{G}\Sigma_{\rm{HI}}\Sigma_{\rm{H_2}} + 2\pi\zeta_{\rm{d}}G\frac{\rho_{\rm{sd}}}{\rho_{\rm{mp}}}\Sigma_{\rm{HI}}^2,
\end{equation}
\citep{Krumholz2013}, where $\zeta_{\rm{d}}$ is taken to be 0.33 and is a dimensionless factor representing the shape of the gas surface density profile, $\Sigma_{\rm{HI}}$ and $\Sigma_{\rm{H_2}}$ are surface densities of \HI and H$_{\rm{2}}$ respectively, $\rho_{\rm{sd}}$ is the mid-plane density of stars and dark matter, $\rho_{\rm{mp}}$ is the volume-weighted mean gas density at the galactic mid-plane. This value is then divided by a factor $\alpha$ in order to find the thermal pressure of the neutral gas at the mid-plane, $P_{\rm{th, mp}}$. This is due to arguments made in \cite{Ostriker2010}, motivated by the fact additional supporting processes will be present at the mid-plane, such as stellar feedback-driven turbulence, magnetic fields and the pressure resulting from thermal gradients caused by cosmic rays. In their paper \cite{Ostriker2010} propose $\alpha = 5$, which is also adopted in \cite{Krumholz2013} and here. Rather than taking a constant value for $\rho_{\rm{sd}}$ as in previous works \citep[e.g.][]{Crain2017}, we build 1D radial profiles of the stellar and dark matter density for each halo, then interpolate the radius of each gas particle to get the average total stellar and dark matter density at its radius. \cite{Diemer2018} investigate the effects of varying this parameter, finding it can range from 0 to 1 M$_{\rm{\odot}}$pc$^{-3}$. Moreover, \cite{Diemer2018} find by varying $\rho_{\rm{sd}}$ the mean H$_2$ masses are similar as those derived using a constant $\rho_{\rm{sd}}$, however the distribution of molecular gas is altered. It is therefore prudent that we do let $\rho_{\rm{sd}}$ vary, given this paper is primarily concerned with the distribution of the cold gas inside galaxies. 

Equation \ref{eqn:disc_pressure} can be re-written using the fact the thermal pressure at the mid-plane can also be expressed in terms of the sound speed of the warm neutral component (WNM), $c_w$;
\begin{equation}
    P_{\rm{th,mp}} = \rho_{\rm{mp}}\tilde{f}_{\rm{w}}c^2_w,
\end{equation}
where $\tilde{f_{\rm{w}}}$ is the ratio of the mass-weighted mean square thermal velocity dispersion to the square of the warm gas sound speed \citep{Krumholz2013}. In this way, the thermal pressure of the gas at the mid-plane can be re-written as
\begin{equation}\label{eqn:Ptherm_midplane}
    P_{\rm{th,mp}} = \frac{\pi{G}\Sigma^2_{\rm{HI}}}{4\alpha}\left\{1 + 2R_{\rm{H_2}} + \left[(1 + 2R_{\rm{H_2}})^2 + \frac{32\zeta_d\alpha\tilde{f}_wc^2_w\rho_{\rm{sd}}}{\pi{G}\Sigma^2_{\rm{HI}}}\right]\right\},
\end{equation}
\citep{Krumholz2013,Ostriker2010}, where $R_{\rm{H_2}} = \Sigma_{\rm{H_2}}/\Sigma_{\rm{HI}}$. This expression for $P_{\rm{th, mp}}$ can be used to form a second constraint on the minimum CNM density (this time labelled $n_{\rm{CNM,hydro}}$);
\begin{equation}\label{eqn:nCNM_hydro}
    n_{\rm{CNM,hydro}} = \frac{P_{\rm{th, mp}}}{1.1k_BT_{\rm{CNM,max}}},
\end{equation}
which must be met in order to ensure hydrostatic equilibrium. Here the factor 1.1 is included to account for the contribution of Helium. Substituting equation \ref{eqn:Ptherm_midplane} into equation \ref{eqn:nCNM_hydro}, \cite{Krumholz2013} then make the further assumption that the H$_{\rm{2}}$ fraction is $\ll 1$ \note{in this regime}, which yields;
\begin{equation}
    n_{\rm{CNM,hydro}} \approx \frac{\pi{G}\Sigma_{\rm{n}}^2}{4\alpha(1.1k_BT_{\rm{CNM,max}})} \times 
    \left[1 + \left(1 + \frac{32\zeta_d\alpha\tilde{f}_wc^2_w\rho_{\rm{sd}}}{\pi{G}\Sigma_{\rm{n}}} \right)^{1/2}\right],
\end{equation}
where $\Sigma_n$ is the surface density of the neutral gas. We adopt a value of $\tilde{f}_w = 0.5$, following \cite{Ostriker2010}, along with an $\alpha = 5$, $T_{\rm{CNM,max}} = 243$K, $\zeta_d = 0.33$ and $c_w = 8$kms$^{-1}$ \citep{Leroy2008}.

Combining equations \ref{eqn:nCNM_2p_simplified} and \ref{eqn:nCNM_hydro}, we arrive at the following condition for the CNM number density;
\begin{equation}\label{eqn:nCNM}
    n_{\rm{CNM}} = \rm max(n_{\rm{CNM,2p}}, n_{\rm{CNM,hydro}}).
\end{equation}
The value obtained for n$_{\rm{CNM}}$ can then be fed into the formalism described in \citet{Krumholz2009} and utilised in \citet{Krumholz2013}. In this analytic prescription, the molecular fraction of neutral gas is dependent on the flux of molecule-dissociating far-UV photons, along with the density of dust grains, on the surface of which H$_{\rm{2}}$ molecules can form. Following this prescription, a dimensionless parameter $\chi$ can be defined, which represents the characteristic optical depth of the dust 
\begin{equation}\label{eqn:chi}
    \chi = 0.72G_0'\left(\frac{n_{\rm{CNM}}}{10\rm{cm}^{-3}}\right),
\end{equation}
\citep{Krumholz2013}. This can then be used to calculate the H$_2$ fraction (and as a consequence the \HI fraction) of the gas, $f_{\rm{H_2}} = \Sigma_{\rm{H_2}}/(\Sigma_{\rm{HI}} + \Sigma_{\rm{H_2}})$, where \cite{McKee2010} showed this can be approximated as
\begin{equation}
f_{\rm{H_2}} \approx 
\begin{cases}
    1 - (3/4)s/(1 + 0.25s),& s < 2\\
    0,              & s\geq 2,
\end{cases}
\end{equation}
where 
\begin{equation}
    s \approx \frac{(1 + 0.6\chi + 0.01\chi^2)}{0.6\tau_c},
\end{equation}
and 
\begin{equation}
    \tau_c = 0.066f_c(Z/Z_{\odot})\Sigma_0.
\end{equation}
$f_c$ is a clumping factor taken to be 5 based on observations by \cite{Krumholz2005} on the same scales as the minimum spatial resolution of EAGLE and $\Sigma_0 = \Sigma_{\rm{n}}/1\, \rm M_{\rm{\odot}}\,\rm pc^{-2}$. 

We note here that our results are robust across different methods of the \HI -- H$_2$ breakdown, as is demonstrated in the appendices (see Appendix \ref{sec:app_gned}). The most significant source of error is the modelling of the galactic feedback processes and chemical enrichment inside the simulation, as will be discussed later in this paper. 

\subsubsection{\HI kinematics}\label{sec:HI_kinematics_method}
\begin{figure}
 \includegraphics[trim=15mm 30mm 15mm 30mm,clip,width=0.95\columnwidth]{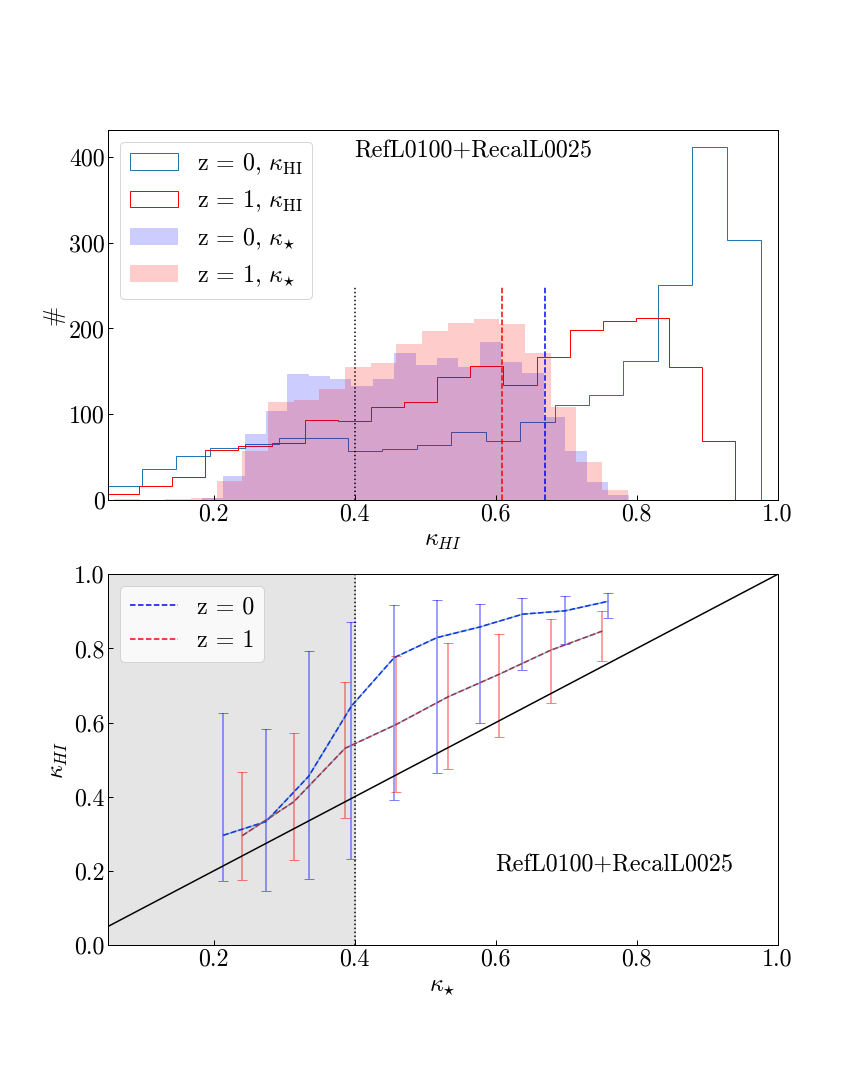}
 \caption{Top plot -- the histogram of $\kappa_{\rm{HI}}$ (computed using an aperture of 30 kpc) at both $z$~$= 1$ (red) and $z$~$= 0$ (blue), combining the resolved galaxies in both the RecalL0025 and RefL0100 simulation boxes to make each histogram. The red dashed and blue dashed lines indicate the mean $\kappa_{\rm{HI}}$ values in our $z$~$= 1$ and $z$~$= 0$ samples respectively. The dotted vertical line indicates the boundary between dispersion-dominated and rotation-dominated stellar kinematics used in \citet{Correa2017}. The shaded blue and red histograms indicates the $z$~$= 0$ and $z$~$= 1$ $\kappa_{\star}$ distributions of our galaxy samples. Lower plot -- $\kappa_{\rm HI}$ versus $\kappa_{\star}$ for our sample of resolved galaxies taken from both the RecalL0025 and RefL0100 simulation boxes, at z~$ = 0$ (blue) and $z$~$= 1$ (red). The 16th -- 84th percentile ranges are indicated by the capped vertical lines. The shaded region indicates the $\kappa_{\star}$ range that is considered to be dispersion-dominated in \citet{Correa2017}. The solid line indicates a one-to-one relationship.}
 \label{fig:kapp_hist}
\end{figure}

In this paper, we investigate the effect $\kappa_{\rm{rot, HI}}$ -- the fraction of the kinetic energy of the \HI gas that is in ordered rotation--has on the DLA's properties. In order to calculate $\kappa_{\rm{rot, HI}}$ we followed the method outlined in previous works exploring stellar rotation \citep[e.g.][]{Sales2010, Correa2017, Thob2019}. Here we used an aperture of 30 kpc in order to calculate the fraction of the total kinetic energy in \HI that is in ordered rotation about the total stellar angular momentum vector, $\hat{z}$, of the galaxy. We did so using the \HI fraction per gas particle f$_{\rm{HI, i}}$;
\begin{equation}
    \kappa_{\rm{rot, HI}} = \frac{\sum^{R < \rm 30 kpc} \frac{1}{2} f_{\rm{HI, i}}m_{\rm{i}} \left[L_{\rm{z, i}}/m_{\rm{i}}R_{\rm i}\right]^2 }{\sum^{R < \rm 30 kpc} \frac{1}{2} f_{\rm{HI, i}}  m_{\rm{i}} v_{\rm{i}}^2}
\end{equation}
\citep[based on equation 1 of ][]{Correa2017}. Here L$_{\rm{z, i}}$ is the angular momentum component of the $i$-th particle parallel to the total stellar angular momentum vector, and $R_{\rm{i}}$ is the projected distance to the stellar angular momentum vector. In this way, $\kappa_{\rm{rot, HI}} =1$ indicates the \HI kinematics in a galaxy is entirely rotation-dominated, while $\kappa_{\rm{rot, HI}} =0$ indicates the \HI gas is dispersion-dominated. In the top panel of Fig.~\ref{fig:kapp_hist} we plot the histogram of $\kappa_{\rm{HI}}$ at both $z$~$= 1$ (red) and $z$~$= 0$ (blue), combining the resolved galaxies in both the RecalL0025 and RefL0100 simulation boxes to make one histogram. We also include the $z$~$= 0$ and $z$~$= 1$ histograms (solid blue/ red histograms respectively) of $\kappa_{\star}$ \report{(the rotational energy fraction of the stars)}, again using both boxes, calculated using the same method as the HI, along with the mean $\kappa_{\rm{HI}}$ values at each redshift. Here we can see the $\kappa_{\rm{HI}}$ distribution is skewed towards 1 at $z$~$= 0$, and to a lesser extent at $z$~$= 1$. We have also plotted a vertical dotted line at $\kappa = 0.4$, which represents the boundary between dispersion-dominated and disc-dominated stellar kinematics used in \citet{Correa2017}. Comparing the $z$~$= 0$ $\kappa_{\rm{HI}}$ and $\kappa_{\star}$ histograms, we can see that $\kappa_{\rm{HI}}$ is more extended, with a mean value of 0.67, as opposed to 0.48 for the stars. This difference is lessened at z~$=$ 1, where the mean values of the \HI and stellar distributions are 0.6 and 0.5, respectively.

We investigate the relationship between stellar and \HI kinematics further in the lower plot of Fig. \ref{fig:kapp_hist}, where we plot $\kappa_{\rm{HI}}$ against $\kappa_{\star}$, again combining the resolved galaxies from the two simulation boxes, RecalL0025 and RefL0100. The grey region highlights the area where $\kappa_{\rm{{\star}}} \leq 0.4$.  The black solid line indicates a one-to-one relationship. The $\kappa_{\rm{HI}}$ values are systematically higher than the corresponding $\kappa_{\star}$ value for a given galaxy at $z$~$= 0$ and $z$~$= 1$, while the $z$~$= 0$ galaxies show a distinctly non-linear relation between the two quantities; with a steep gradient between $\kappa_{\star} = 0.2 - 0.5$ and a shallower gradient beyond this. Since there is no clear transition point between two kinematic populations of galaxies at $z$~$= 1$ (either in the \HI or stars) we use the mean of the $\kappa_{\rm{HI}}$ values of our sample, 0.6, to be our transition point between dispersion-dominated ($\kappa_{\rm{HI}} < 0.6$) and rotation-dominated ($\kappa_{\rm{HI}} \geq 0.6$) \HI kinematics. At $z$~$= 0$ we instead use the higher threshold of $\kappa_{\rm{HI}} \geq 0.67$ for $z$~$= 0$, given this is the mean value of the distribution.

We explored the effect of altering the aperture by re-plotting Fig.~\ref{fig:kapp_hist}, with an aperture of 50 kpc for the \HI instead (not shown for brevity). We found that increasing the aperture decreases the mean of the $z$~$= 0$ and $z$~$= 1$ $\kappa_{\rm{HI}}$ distributions to $\geq$ 0.55, while the $z$~$= 1$ $\kappa_{\rm{HI}}$ -- $\kappa_{\star}$ relation moves away from the 1:1 line towards higher $\kappa_{\rm{HI}}$ values for a given $\kappa_{\star}$. Despite these differences, the overall trends are similar -- the $z$~$= 0$ $\kappa_{\rm{HI}}$ distribution showed a significant skew towards higher $\kappa_{\rm{HI}}$ values when compared with the $z$~$= 1$ distribution, while the $\kappa_{\rm{HI}}$ value of galaxies increased with $\kappa_{\rm{\star}}$ at both redshifts using both apertures, with the \HI showing a significantly higher rotation to total kinetic energy ratio than the stars. 
\subsubsection{Measurements of DLAs}\label{sec:DLA_method}
We use three different methods to estimate the distribution of high column density gas associated with each galaxy. The first two involve taking the covering fraction, f$_{\rm{cov}}$, of each galaxy. To do so, we orient the galaxy face-on (using its total stellar angular momentum axis) and use SPH interpolation to project it onto a 2D grid. \report{In our fiducial method, based on \report{\citet{Nagamine2004}}, the grid has a length of 2~R$_{\rm{200c}}$ (centered on the minimum of the potential well of the galaxy) and is made up of $n^2$ grid cells}. Here $n$ was chosen to ensure a cell size of 3 kpc (approximately four times the softening length, 0.7~pkpc). We also varied this from 2 kpc to 5 kpc, finding this had an insignificant effect on our results. We then find the \HI column density (N$_{\rm{HI}}$) associated with each 2D grid cell by summing over all particles that meet three spatial criteria: an x-position inside the defined cell, a y-position in the defined cell and a z-position within 2 R$_{\rm{200c}}$;
\begin{equation}
    N_{\rm{HI}} =  \frac{L_{\rm{cell}}}{m_p} \sum_{\rm{i}} \rho_{\rm{HI, i}},
\end{equation}
where $m_p$ is the mass of a proton, $L_{\rm{cell}}$ is the length of each grid cell and $\rho_{\rm{HI}}$ is the SPH interpolated \HI density of each cell. f$_{\rm{cov}}$ is then defined as the fraction of the total number of grid cells with N$_{\rm{HI}} >$ 10$^{20.3}$ cm$^{-2}$. A second covering fraction, f$_{\rm{cov, 70 kpc}}$ was also defined using a fixed grid length of 140~kpc, independent of the size/ mass of the system. 

Finally, volume filling fractions, f$_{\rm{vol}}$, were also computed using a 3D grid containing $n^3$ cells and a grid size of 2~R$_{\rm{200c}}$. This time the column density per cell was defined as $(\rho_{\rm{HI}}/ m_p) L_{\rm{cell}}$ and the covering fraction was calculated as the number of cells with N$_{\rm{HI}} > 10^{20.3}$~cm$^{-2}$ divided by the total number of cells. The volume filling fractions allow us to gauge the 3D distribution of the high column density \HI about a galaxy, for example should a galaxy have a high volume filling fraction and a low covering fraction, this would imply a highly disturbed DLA distribution. Moreover, the individual cells used for the volume filling factors allow us to understand the local properties of individual DLAs in a way that is not possible averaging over a column. 

\begin{figure}
 \includegraphics[trim=0mm 2mm 0mm 0mm,clip,width=0.95\columnwidth]{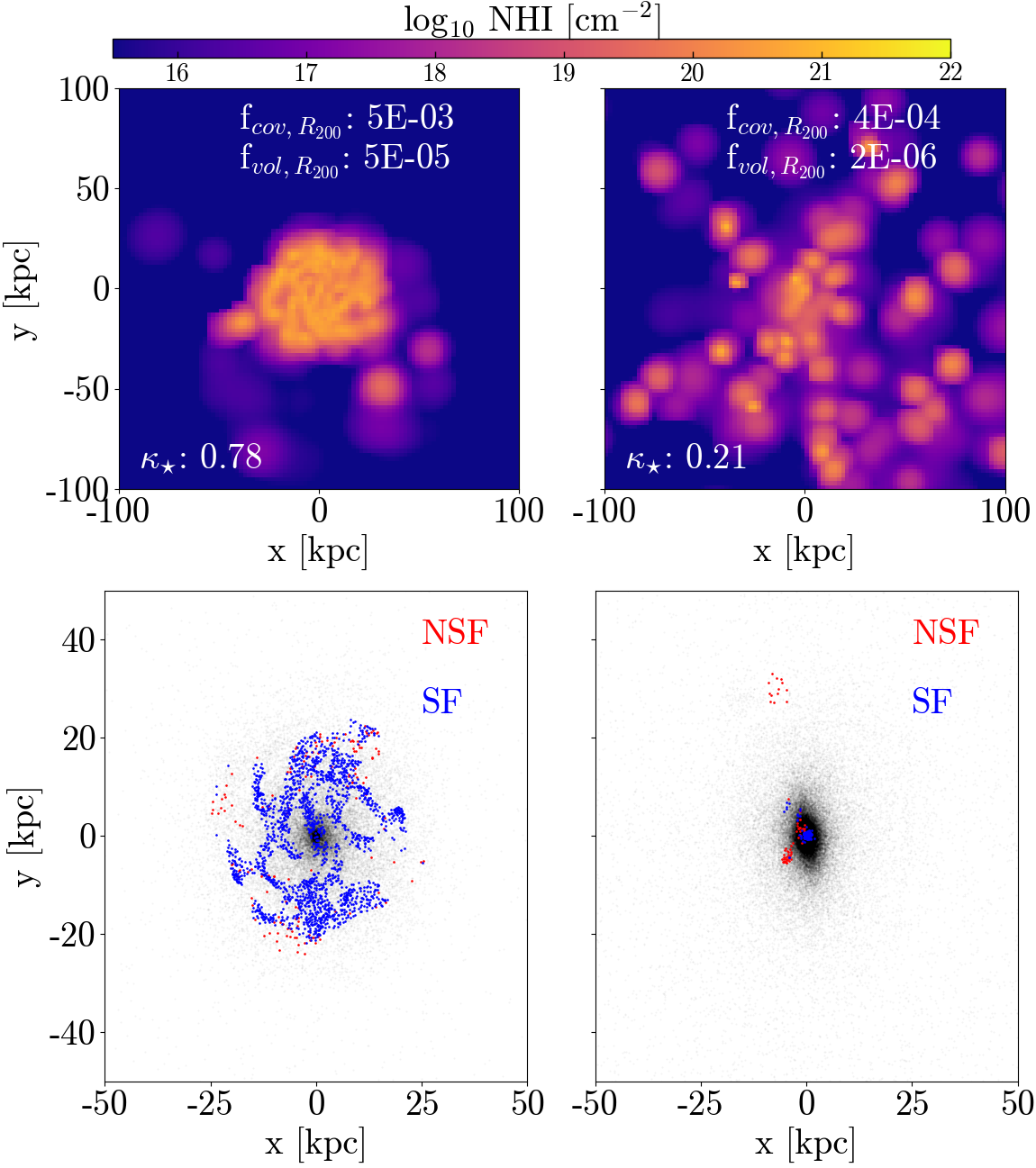}
 \caption{Upper row -- 2D gridding of two galaxies taken from the EAGLE RefL0100 simulation at$z$~$= 0$. The galaxy on the right has a low stellar rotation-to-dispersion kinematic ratio ($\kappa_{\star}$), while the galaxy on the left is rotation-dominated. Plots are orientated face-on. The 2D face-on DLA covering fraction (f$_{\rm{cov, R_{200}}}$) and 3D volume filling fraction for DLAs (f$_{\rm{vol, R_{200}}}$), calculated using an aperture equal to virial radius (R$_{\rm{200c}}$) of each galaxy, are shown on each plot, along with the $\kappa_{\star}$ values. Lower row -- the application of the \citet{Mitchell2018} ISM selection criteria to the same two galaxies, zooming in to the central 50~kpc of each galaxy. The star-forming (SF) ISM gas particles are shown in blue, while the non-star-forming (NSF) particles are in red. Black points represent the position of star particles.}
 \label{fig:NHI_ex}
\end{figure}

To illustrate our method, we plotted two examples of the 2D N$_{\rm{HI}}$ grids obtained using a total grid side length of 200~kpc for two galaxies extracted from the EAGLE RefL0100 simulation at $z$~$= 0$ (upper plots of Fig.~\ref{fig:NHI_ex}). The left plot was obtained for a galaxy with dispersion-dominated stellar kinematics, while the lower plot was obtained for a galaxy with a high stellar rotation-to-dispersion kinematic ratio. As expected the late-type galaxy has higher covering and volume filling factors, and, due to projections, the covering fraction is usually higher than the volume filling fraction.

\subsection{ISM versus CGM}\label{sec:cgm_method}
We further split the grid cells into those containing gas associated with the ISM and those that do not, categorising the latter as CGM sightlines and the former as ISM sightlines. To class individual particles as being exclusively part of the CGM or ISM, we used the method outlined in \citet{Mitchell2018}. This method computes whether or not a gas particle is rotationally supported, along with its temperature and position within the galaxy. To be in the ISM, the particle has to satisfy the following criteria;
\begin{itemize}
    \item The temperature of the gas particle must be below 10$^{5}$~K, unless the hydrogen number density, n$_{\rm{H}}$, is greater than 500~cm$^{-3}$.  
    \item The gas must be dense, with n$_{\rm{H}} >$ 0.03 cm$^{-3}$, or be rotationally supported. The latter requires the gas to satisfy the following numerical criteria:
    \begin{equation}
        - 0.2 < {\rm log_{\rm{10}}} \left(\frac{2 \epsilon_{\rm{k, rot}}}{\epsilon_{\rm{grav}}}\right) < 0.2,
    \end{equation}
    and
     \begin{equation}
        \frac{\epsilon_{\rm{k, rot}}}{\epsilon_{\rm{k, rad}} + \eta_{\rm{th}}} > 2,
    \end{equation}
    where $\epsilon_{\rm{k, rot}}$ is the specific kinetic energy of the particle that is attributed to rotation about the galactic centre, $\epsilon_{\rm{k, rad}}$ is the same but for radial motion, $\epsilon_{\rm{grav}}$ is the specific gravitational energy of the particle ($GM(r)/r$), and $\epsilon_{\rm{th}}$ is the specific internal energy of the particle.
    \item The radial distance of the gas particle to the galaxy centre must be below 0.5 R$_{\rm{200c}}$.
\end{itemize}
Furthermore, following \citet{Mitchell2018}, once the ISM particles have been selected, any gas particles at a radius greater than r$_{\rm{90}}$ (the radius enclosing 90$\%$ of the ISM mass) are considered CGM instead, while if the galaxy has a M$_{\rm{ISM}}$/M$_{\rm{\star}}$ ratio of less than 0.1, any gas particles beyond 5 r$_{\rm{1/2}}$ are also considered to be CGM. 
 
The results of applying the above set of criteria to the galaxies of the top panels of Fig. \ref{fig:NHI_ex} are shown on the bottom row. \note{Here we have zoomed in to the central 50 kpc region of each galaxy; the star-forming ISM particles are plotted in blue and the non-star-forming ISM particles are coloured red.} The star particles are shown as black points in the background. 

These methods allow an in-depth exploration of the distribution of \HI inside a galaxy. In particular, we compare high column density \HI in the ISM with the CGM. 

\section{Results}\label{sec:results}
\subsection{Mass scaling relations and evolution with redshift}\label{sec:HI_mass}
\begin{figure}
 \includegraphics[trim=0mm 9mm 0mm 5mm,clip,width=0.95\columnwidth]{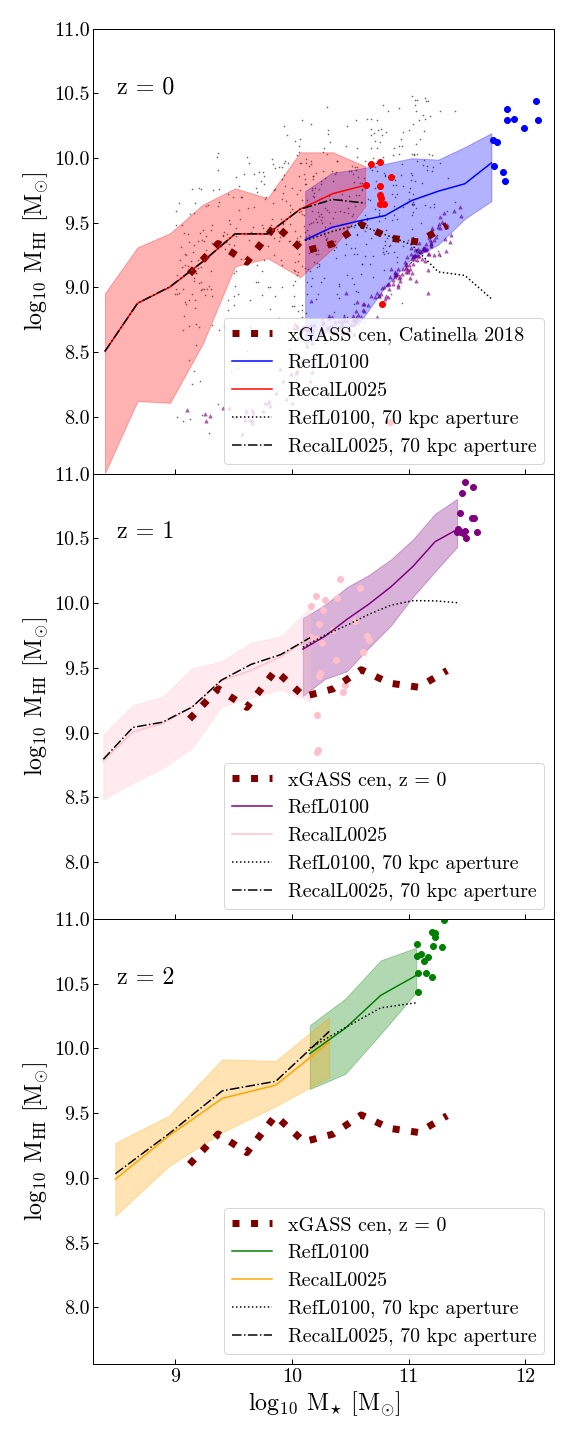}
 \caption{The M$_{\rm{HI}}$--M$_{\rm{\star}}$ relation at $z$~$= 0$ (top row), $z$~$= 1$ (middle row) and $z$~$= 2$ (bottom row), as calculated for both EAGLE box RecalL0025 and RefL0100. The solid lines indicate the median values in 10 stellar mass bins for each galaxy sample, while the shaded regions are the 16th -- 84th percentile range inside each stellar mass bin. Points indicate galaxies in stellar mass bins containing fewer than 10 galaxies. Also plotted is a representative sample of the M$_{\rm{HI}}$ and M$_{\rm{\star}}$ values obtained by the xGASS survey \citep{Catinella2018} (black dots/purple triangles for detections/ non-detections), along with the median values for central galaxies (dashed line). We over-plot the M$_{\rm{HI}}$--M$_{\rm{\star}}$ relation obtained using a 70 kpc aperture for the \HI gas for both the RefL0100 run (black, dotted line) and the RecalL0025 run (black, dot-dashed line).}
 \label{fig:MHI_Mstar}
\end{figure}

Initially we investigate the redshift evolution of the global \HI budget of haloes in order to understand the typical \HI properties of galaxies and their surroundings in the redshift range probed by FLASH. In particular, we are interested in the spatial distribution/ kinematics of the \HI in these galaxies and to understand what a typical \HI mass-selected galaxy looks like at $z$~$= 1$.  Fig.~\ref{fig:MHI_Mstar} shows the M$_{\rm{HI}}$--M$_{\rm{\star}}$ relation for both the RefL0100 and RecalL0025 galaxy samples at varying redshift. Also plotted are the estimates of \citet{Catinella2018}, from targeted \HI observations of a stellar-mass selected sample at $z\approx 0$. This survey comprises of 1179 galaxies in the local Universe. We compute the median M$_{\rm{HI}} -$ M$_{\rm{\star}}$ relation for a representative sample\footnote{\href{https://xgass.icrar.org/data.html}{https://xgass.icrar.org/data.html}} of the xGASS central galaxies (dashed lines). Here we initially calculated the \HI mass using all gas particles attributed to the halo using SUBFIND (our methods are explored later in the section). At $z$~$= 0$, the \HI masses obtained using the larger box are systematically lower than those obtained at the same stellar mass for the higher resolution box. On further analysis we found there was no significant difference in the median cold gas fractions measured for both simulation boxes at $z$~$= 0$, the difference in \HI masses is instead driven by the relative abundances of H$_2$ and \HI. As was also seen in \citet{Lagos2015}, the galaxies inside the RefL0100 EAGLE simulation have systematically higher ISM mean metallicities (a fact that will also become important when considering the properties of associated DLAs, see Section \ref{sec:DLA_props}) and since the post-processing \HI prescription relies on dust as a catalyst of H$_2$ formation, which we explicitly link to metallicity (see Section \ref{sec:HI_method}), this results in higher M$_{\rm{H_2}}$/ M$_{\rm{neut}}$ ratios \citep[where M$_{\rm{neut}}$ is neutral gas mass; for a discussion on this see section 4.1 of ][]{Lagos2015}. This effect is present both a $z$~$= 0$ and $z$~$= 1$, but is more significant at lower redshift. 

\begin{figure}
 \includegraphics[width=0.95\columnwidth]{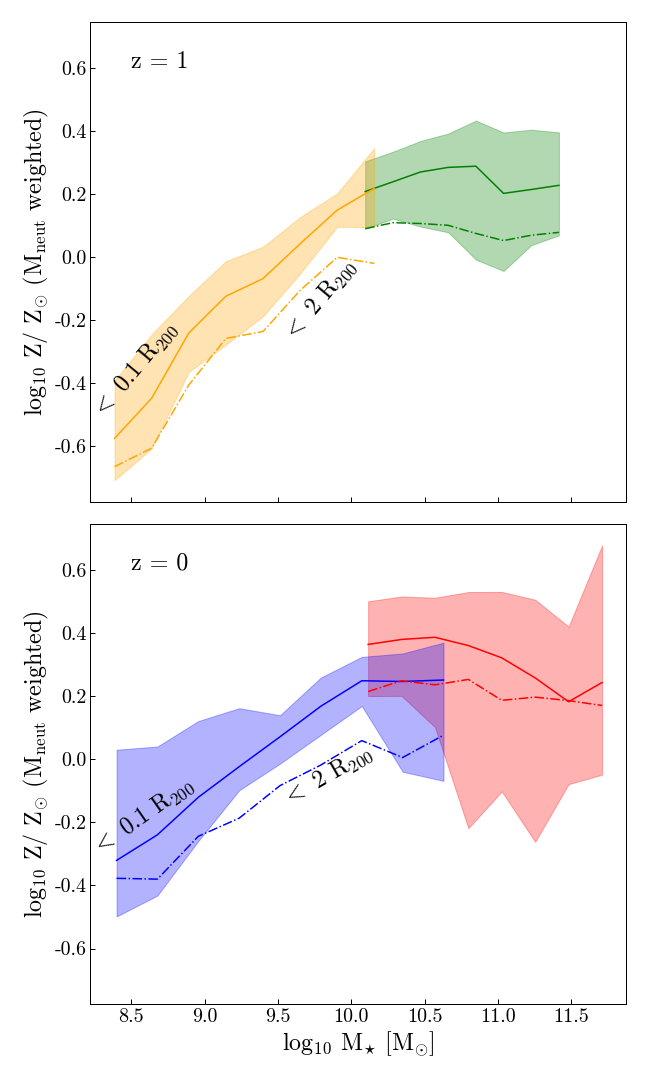}
 \caption{The mean metallicity, weighted by the neutral gas mass M$_{\rm{neut}}$, calculated for all gas below 2 radii; 0.1 R$_{\rm{200c}}$ (solid lines) and 2 R$_{\rm{200c}}$ (dot-dashed lines) as a function of stellar mass, for galaxies in the RefL0100 (blue/ green) and RecalL0025 (red/ orange) EAGLE simulations at $z$~$= 1$ (upper plot) and $z$~$= 0$ (lower plot). The shaded areas show the 84th -- 16th percentile range.}
 \label{fig:Z_rad}
\end{figure}

We also investigated how the relative metallicity of the \HI gas in the galaxies changes with radius in the RefL0100 and RecalL0025 simulations, in order to further understand the disparity seen in the \HI masses between the different boxes at $z$~$= 0$, along with why this disparity disappears at higher redshift. We measured the mean neutral-mass-weighted metallicity (Z$_{\rm{{neut}}}$) of all gas below 2 different radii; 0.1 R$_{\rm{200c}}$ and 2 R$_{\rm{200c}}$. We plot this for $z$~$= 1$ and $z$~$= 0$ in Fig.~\ref{fig:Z_rad}. At both redshifts, the disagreement between the mean metallicities in the two EAGLE simulations is more pronounced when averaging over the entire galaxy. This disagreement is also seen when averaging over all gas particles below 0.1~R$_{\rm{200c}}$ for $z$~$= 0$. On the other hand, at $z$~$= 1$, the two resolution boxes show a good agreement for the mean neutral-mass-weighted metallicities below 0.1~R$_{\rm{200c}}$ for galaxies in the same stellar mass bin.

The better agreement between metallicities in the inner parts of the galaxy, where H$_2$ is primarily expected to form, results in the better agreement between H$_{\rm{2}}$ -- \HI mass fractions in the centre of galaxies at $z$~$= 1$. Furthermore, the systematic increase in the metallicity of the neutral gas in the outer parts of the galaxy, or the CGM, does not have a significant impact on the molecular/ atomic decomposition, since this gas is at temperatures and pressures that prohibit the formation of H$_2$. We also see this result when comparing the SF (star-forming) and NSF (non-star-forming) gas metallicities between the two EAGLE boxes, finding the agreement between the Z$_{\rm{NSF}}$ values is better at low z, while the opposite is true for the star-forming gas (we have not included this plot for brevity). 

Focussing again on Fig. \ref{fig:MHI_Mstar}, our fiducial $z$~$= 0$ RefL0100 results do not show the downturn seen in the M$_{\rm{HI}}$--M$_{\rm{\star}}$ relation at $z$~$= 0$ by \citet{Crain2017} also using EAGLE. This is due to the fact the latter utilise a 70-kpc aperture, while we consider all gas associated with the galaxy via SUBFIND when determining \HI masses. According to the \HI size--mass relation (known to be extremely robust, see \citet{Stevens2019b} and references therein), we would expect galaxies with \HI masses above 10$^{10.8}$ M$_{\odot}$ to have a significant fraction of their \HI extending beyond 70 kpc and indeed we do see that the use of the 70-kpc aperture has the largest impact at high M$_{\star}$/ M$_{\rm{HI}}$. However, Fig. \ref{fig:MHI_Mstar} shows a significant reduction in the \HI mass measured using a 70-kpc aperture at \HI masses of $>$ 10$^9$ M$_{\odot}$ (or alternatively M$_{\star}$ $>$ 10$^{10}$ M$_{\odot}$) at $z$~$= 0$. This discrepancy in \HI mass is seen at all redshifts (however is greatest at $z$~$= 0$) and implies a significant mass of \HI lies outside the galactic disc (i.e. in the CGM) of systems with M$_{\star}$ $>$ 10$^{10}$ M$_{\odot}$ at $z$~$= 0$, or M$_{\star}$ $>$ 10$^{10.5}$ M$_{\odot}$ at $z$~$= 1$.

\begin{figure}
 \includegraphics[trim=4mm 5mm 4mm 0mm,clip,width=0.95\columnwidth]{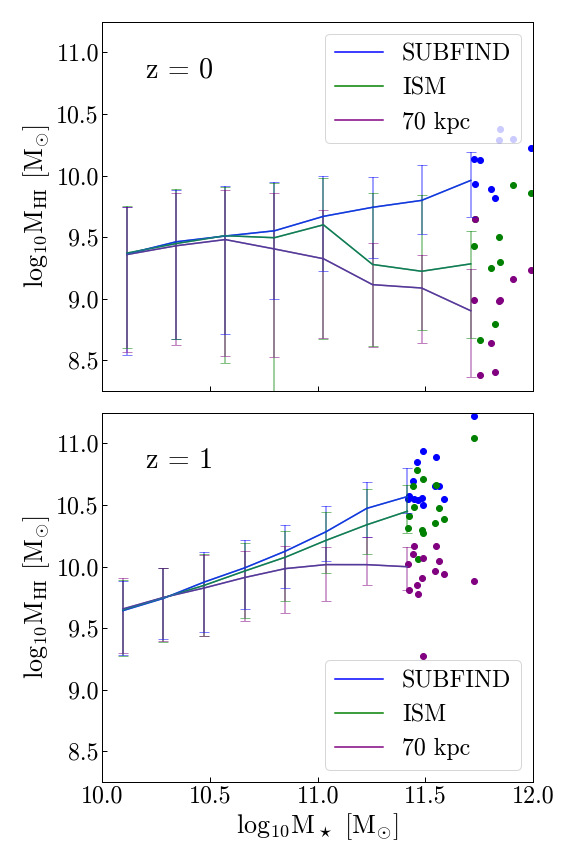}
 \caption{The M$_{\rm{HI}}$--M$_{\rm{\star}}$ relation at $z$~$= 0$ (upper plot) and $z$~$= 1$ (lower plot) for the RefL0100 galaxy sample, with three different methods of calculating the \HI mass indicated. `SUBFIND' indicates all gas particles attributed to a halo are used to calculate the \HI mass, while `70 kpc' indicates all gas particles within a 70 kpc radius and `ISM' uses just the \HI mass attributed to the ISM using the method described in Section \ref{sec:cgm_method}. The values for stellar mass bins that contain less than 10 galaxies are shown as points, while the  16th -- 84th percentiles are indicated by the vertical lines.}
 \label{fig:HI_app}
\end{figure}

When we plot the ISM \HI masses computed as detailed above, we see the \HI masses are reduced substantially, particularly at $z$~$= 0$. This is demonstrated in Fig. \ref{fig:HI_app} where we show the M$_{\rm{HI}}$-M$_{\rm{\star}}$ relation at $z$~$= 0$ and $z$~$= 1$ plotted for the RefL0100 galaxy sample using different methods of calculating the \HI mass. At both redshifts, using just the \HI associated with the ISM reduces the \HI masses calculated (albeit to a lesser extent at $z$~$= 1$) indicating that a significant mass of \HI is in the CGM of higher-mass galaxies at $z$~$= 0$. 

We find the M$_{\rm{HI}} -$ M$_{\rm{\star}}$ relation shows an increase in normalisation in both the reference and re-calibrated box as redshift increases (more noticeable for the RefL0100 run). This can be seen clearly by comparing the offset of the median M$_{\rm{HI}} -$ M$_{\rm{\star}}$ profile (solid lines) to the median of the xGASS galaxies (dashed lines) in Fig.~\ref{fig:MHI_Mstar}. The offset from the $z$~$= 0$ M$_{\rm{HI}}$--M$_{\rm{\star}}$ relation increases with M$_{\rm{\star}}$ in both simulation boxes.

Both RefL0100 and RecalL0025 agree well with the observations at $z$~$= 0$ within the stellar mass range adopted here. This gives us confidence that we can use EAGLE to explore the ISM and CGM \HI content of galaxies and halos.

\subsection{The covering fraction of DLAs as a function of galaxy properties}
Now that we have explored the global \HI scaling relations, and in particular inferred the presence of a significant \HI mass outside the galactic disc of high mass galaxies, we will dive into the key results of the paper: the redshift evolution of DLA properties, along with the properties of their host haloes. We begin by investigating the distribution of DLAs inside galaxies using the 2D covering fractions (f$_{\rm{cov, R_{\rm{200}}}}$; for details on how these were calculated refer to Section \ref{sec:DLA_method}).

\begin{figure}
 \includegraphics[trim=2mm 5mm 1mm 0mm,clip,width=0.95\columnwidth]{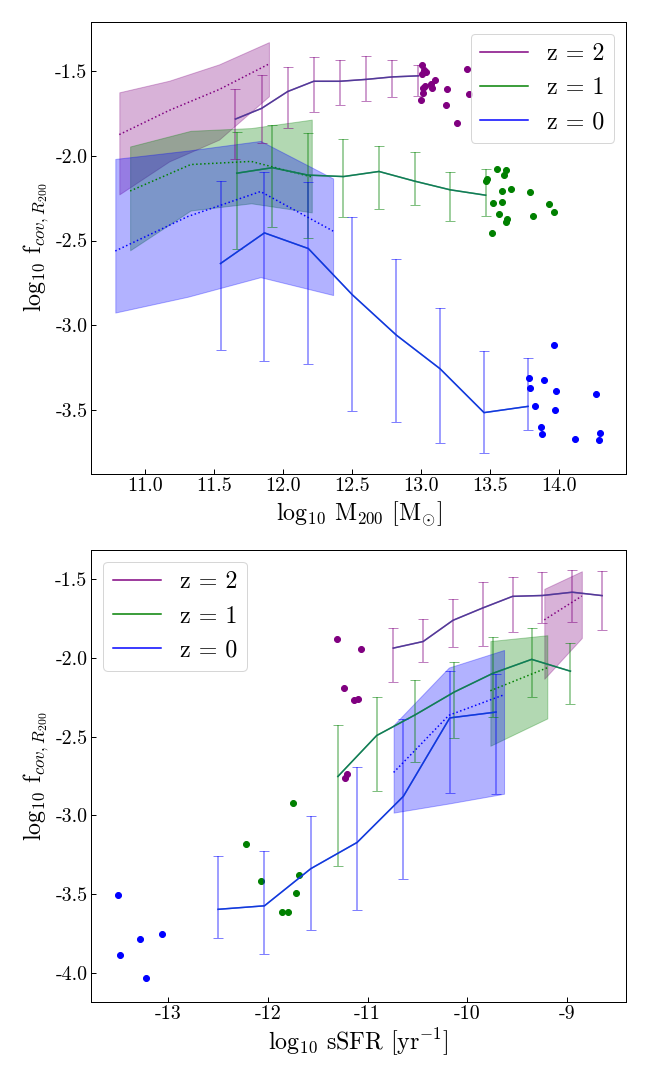}
 \caption{The DLA covering fraction as a function of M$_{\rm{200}}$ (top plot), sSFR (defined as SFR/ M$_{\rm{\star}}$, lower plot) at $z$~$= 2$ (purple), $z$~$= 1$ (green) and $z$~$= 0$ (blue). The median values for the sample of galaxies taken from the EAGLE RefL0100 simulation are shown using the solid lines, while those from the RecalL0025 are shown using dotted lines. The 84th -- 16th percentile range is indicated by the vertical lines for the RefL0100 simulation and the shaded area for the RecalL0025 galaxies. Where bins contain fewer than 10 galaxies, individual galaxies are plotted as points.}
 \label{fig:fcov_prop}
\end{figure}

The top panel of Fig.~\ref{fig:fcov_prop} shows f$_{\rm{cov, R_{\rm{200}}}}$ as a function of M$_{\rm{200}}$ for the EAGLE RefL0100 (solid lines) and the RecalL0025 (dotted lines) simulations. Both boxes show a redshift evolution in the median covering fraction, with higher values at higher redshift. 

Focussing on $z$~$= 0$, we can see that between 10$^{11.5}$ M$_{\rm{\odot}} < $M$_{\rm{200}} \leq$ 10$^{12}$ M$_{\rm{\odot}}$,  f$_{\rm{cov, R_{\rm{200}}}}$ increases with halo mass, but there is a turnover at $\sim$ 10$^{12.25}$ M$_{\rm{\odot}}$. \note{This turnover is present in both simulation boxes and in the same halo mass range. On the other hand, at $z$~$= 1$, f$_{\rm{cov, R_{\rm{200}}}}$ is close to flat and independent of M$_{200}$.}

\note{We have also included the$z$~$= 2$ values in Fig. \ref{fig:fcov_prop} in order to ascertain whether the f$_{\rm{cov, R_{\rm{200}}}}$ -- M$_{\rm{200}}$ trends remains flat at higher redshift. Instead, we see the f$_{\rm{cov, R_{\rm{200}}}}$ -- M$_{\rm{200}}$ relation is inverted at $z$~$= 2$ compared with $z$~$= 0$; f$_{\rm{cov, R_{\rm{200}}}}$ increases with increasing M$_{\rm{200}}$ in both the RefL0100 and RecalL0025 EAGLE simulations, albeit weakly}. Overall, these results indicate a complex, non-linear relationship between halo mass and DLA covering fractions, with the covering fractions of high-mass (M$_{\star} > 10^{11}$ M$_{\odot}$) galaxies becoming more significant at higher redshift. This is most likely due to the fact the vast majority of galaxies at $z$~$= 2$ are actively forming stars, as opposed to $z$~$= 1$, where galaxies at higher M$_{\rm{200}}$ become more passive.

The position of the turnover at $z$~$= 0$ corresponds to the turnover in the GSMF seen both observationally \citep[e.g.][]{Davidzon2017, Tomczak2014} and theoretically \citep[e.g.][]{Beckmann2017} when AGN feedback is accounted for. This \report{could imply that AGN feedback is acting to reduce the \HI with DLA-like column densities into a lower column density phase. Or alternatively, the \report{inefficiency of stellar feedback at driving gaseous outflows in high-stellar-mass galaxies is also likely to reduce the \HI covering fraction.} Furthermore, this turnover could also signify an increase in the clumpiness, or \HI-mass-weighted density, of DLAs (we discuss this further in Section \ref{sec:fcov_vs_fvol}).}

\begin{figure}
 \includegraphics[trim=2mm 5mm 23mm 0mm,clip,width=0.95\columnwidth]{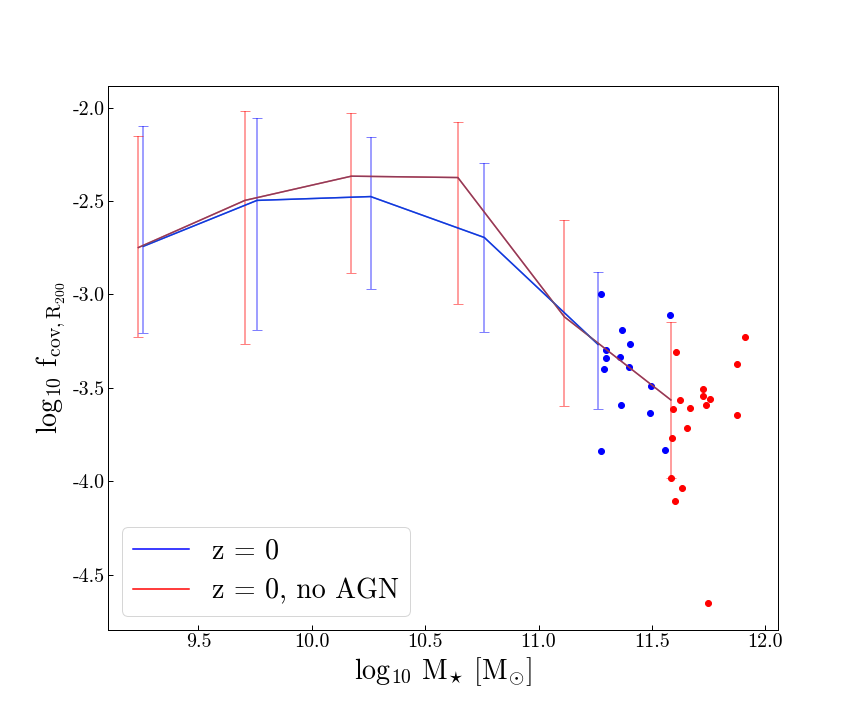}
 \caption{\report{The DLA covering fraction as a function of M$_{{\star}}$ at $z$~$= 0$. The median values for the sample of galaxies taken from the EAGLE RefL0050N0752 simulation are shown using the blue solid lines, while those from the NoAGNL0050N0752 are shown in red. The 84th -- 16th percentile range is indicated by the vertical lines. Where bins contain fewer than 10 galaxies, individual galaxies are plotted as points.}}
 \label{fig:fcov_prop_agn}
\end{figure}

\report{We test the impact of AGN feedback on the trends in the DLA covering fraction -- M$_{200}$ relation seen in Fig. \ref{fig:fcov_prop} by re-plotting the $z$~$= 0$ results for an EAGLE simulation with/without AGN feedback included; RefL0050N0752 (in blue) and NoAGNL0050N0752 (in red) respectively, controlling for stellar mass instead of halo mass (Fig. \ref{fig:fcov_prop_agn}). Here we see both simulations show a downturn in the DLA covering fraction at high-stellar-masses, but this occurs at a stellar mass that is $\sim$ 0.5 dex higher when AGN feedback is not included. This supports our conclusion that AGN feedback is acting to reduce the DLA covering fraction of high-mass galaxies. Although, this effect becomes insignificant beyond a stellar mass of $\sim$ 10$^{11}$ M$_{\odot}$.}

\begin{figure}
 \includegraphics[trim=0mm 0mm 0mm 0mm,clip,width=0.95\columnwidth]{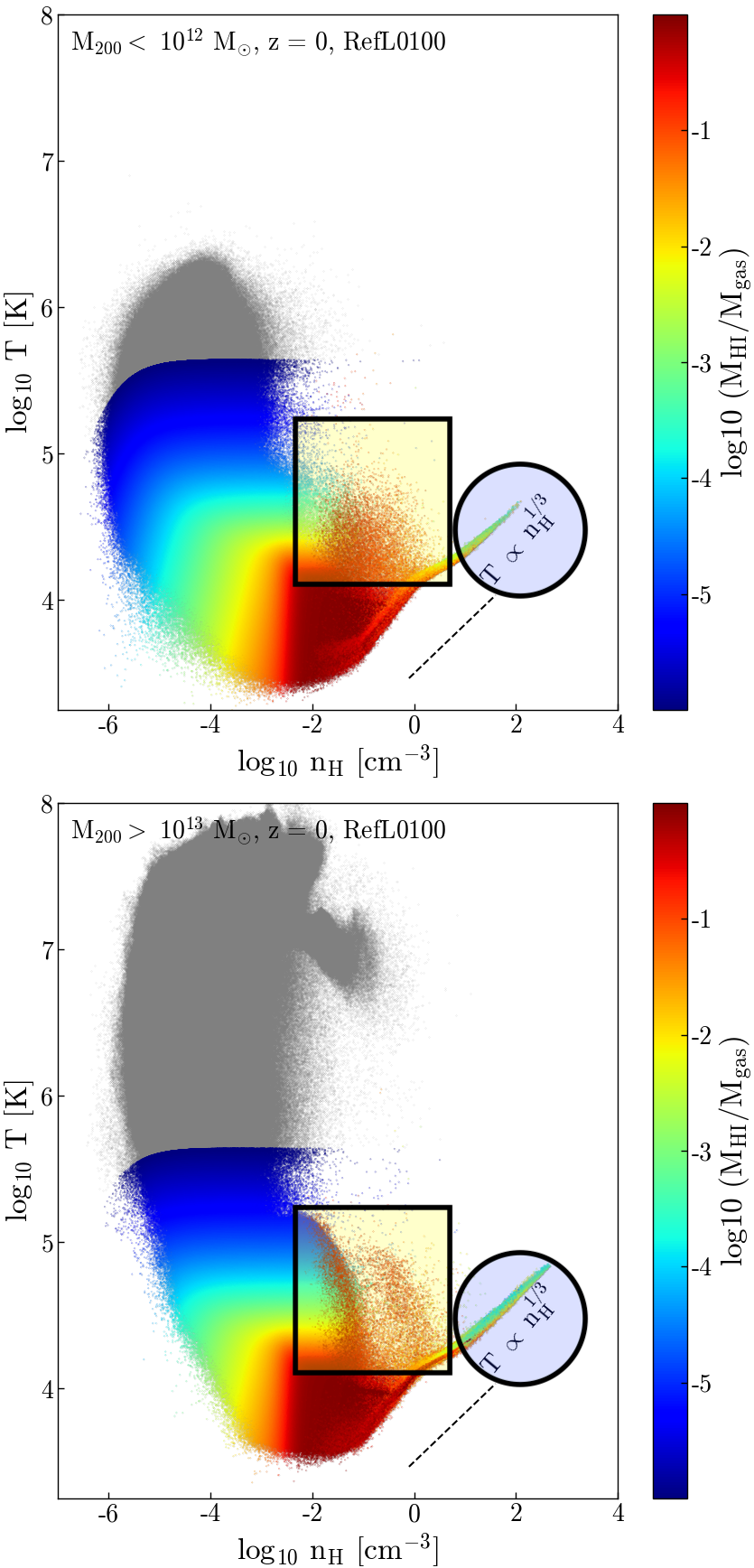}
 \caption{Temperature versus hydrogen number density (n$_{\rm{H}}$) plots for the gas particles, coloured by \HI mass fraction, in galaxies of halo mass $<$ 10$^{12}$ M$_{\rm{\odot}}$ (top plot) $>$ 10$^{13.5}$ M$_{\rm{\odot}}$ (lower plot) taken from the RefL0100 simulation at $z$~$= 0$. All other gas particles associated with the galaxies are included as grey points. The rectangles and circles highlight areas of interest. The T $\propto$ n$_{\rm{H}}^{1/3}$ line indicates the equation of state imposed in EAGLE to avoid artificial fragmentation below the resolution limit.}
 \label{fig:gas_phase}
\end{figure}

\note{It} is also possible the transition of the high density \HI associated with DLAs into molecular hydrogen is more efficient in systems with halo mass greater than 10$^{12}$ M$_{\odot}$, while the cooling of the cold gas is less efficient. We plot the gas phase diagram of the gas particles in galaxies with M$_{\rm{200}} >$ 10$^{13}$ M$_{\rm{\odot}}$ and M$_{\rm{200}} <$ 10$^{12}$ M$_{\rm{\odot}}$ in Fig.~\ref{fig:gas_phase}, colouring the particles according to their \HI gas mass fraction (M$_{\rm{HI}}$/M$_{\rm{gas}}$). Here we see the characteristic equation of state (T $\propto \rm{n_{H}}^{1/3}$; applied to avoid artificial fragmentation as a power law) in the lower right of each panel. It is this gas that will be star-forming. Concentrating on the particles with the highest \HI mass fraction in red, the shaded rectangle highlights the fact the high halo mass galaxies have a large mass of \HI in the temperature range of 10$^4$ K to 10$^5$ K, at intermediate density (n$_{\rm{H}} = 10^{-2}$ -- 10$^{0}$ cm$^{-3}$). This indicates an additional heating source of the diffuse \HI in the latter case, such as AGN. 

\note{Additionally, in the lower right of the plot (in the area marked by the shaded blue circle) we can see the \HI fraction in star-forming gas (i.e. the gas lying in the imposed equation of state) is significantly higher in the bottom panel compared to the top panel, indicating that while there is an effective source of heating in the higher mass galaxies, there is also a significant build up of high-density, star-forming gas}. This heating/star-forming gas build-up is in line with results from \citet{Bower2017}, who ascribed it to the suppression of star formation-driven outflows in the hot corona of massive galaxies, leading to a build up of gas in the central regions of the galaxy and a subsequent period of non-linear black hole growth and intense AGN feedback. This then heats the corona and prevents further cold gas accretion onto these galaxies.

Focussing on what these results mean for FLASH, the trends in covering fraction seen here link directly to both the probability of high/ low impact parameters for given galaxies, along with the probability of an intervening absorption detection should each galaxy be situated along a given sightline. The higher covering fractions we see at $z$~$= 1$ link to the recent finding of a high impact parameter absorption detection in an early-type galaxy at redshift z~$= 0.3562$ in \citet{Allison2020}. Our results suggest higher impact parameter (relative to the size of the galaxy halo) absorption detections are more likely the further you look back in redshift.

There is a clear positive correlation between the DLA covering fraction and sSFR (lower panel of Fig. \ref{fig:fcov_prop}) and H$_{\rm{2}}$ mass fraction (calculated as M$_{\rm{H_2}}$/ M$_{\rm{200}}$, not shown in figure for brevity) at $z$~$= 0$, $z$~$= 1$ and $z$~$= 2$, \note{indicating that galaxies with higher sSFR also have a more extended distribution of DLAs, relative to R$_{\rm{200c}}$.} This correlation is steepest at $z$~$= 0$ and the gradient is reduced at higher redshift. Consequently, the higher redshift DLAs in EAGLE are attributed to galaxies with a wider range of sSFRs at $z$~$= 1$ and $z$~$= 2$. This hints that \HI absorption surveys such as FLASH will be tracing a representative sample of galaxies at $z$~$= 1$ (although this is contingent on the CNM covering fraction following the same trends as that of the DLAs).

\begin{figure}
 \includegraphics[trim=3mm 30mm 15mm 40mm,clip,width=0.95\columnwidth]{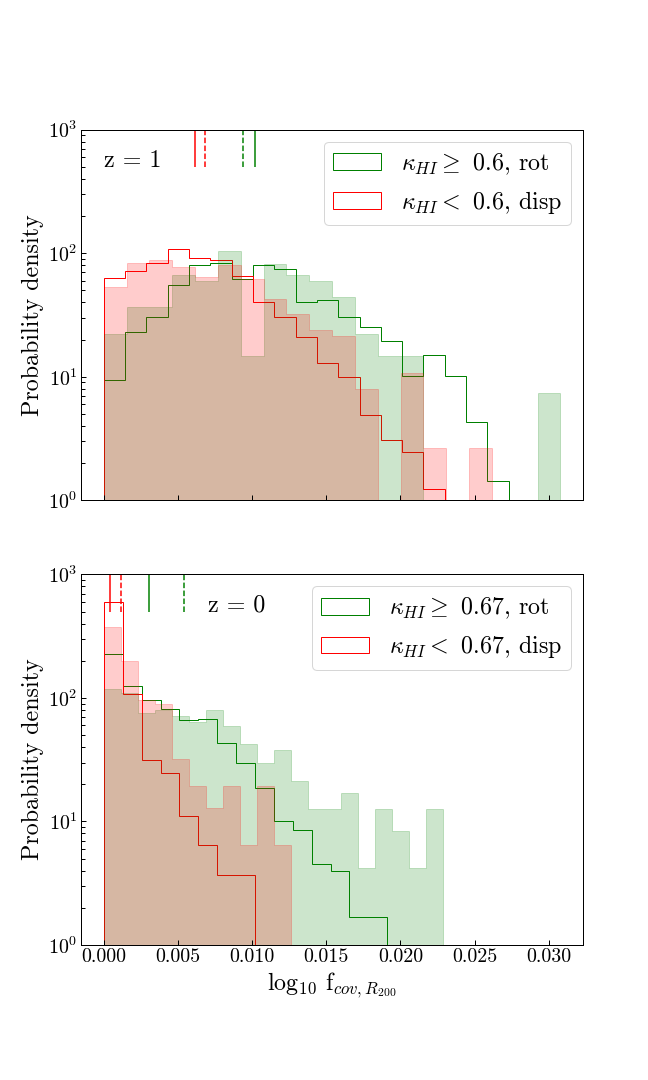}
 \caption{The probability density distribution of f$_{\rm{cov, R_{\rm{200}}}}$ values for galaxies taken from RefL0100 (step histograms) and RecalL0025 (shaded histograms) at $z$~$= 0$ (lower plot) and $z$~$= 1$ (upper plot), then split into two bins according to their $\kappa_{\rm{HI}}$ values, with galaxies with rotation-dominated \HI kinematic morphologies in green and those with dispersion-dominated \HI kinematics in red (following the categorisation detailed in Section \ref{sec:HI_kinematics_method}). The median f$_{\rm{cov, R_{\rm{200}}}}$ values for the RefL0100 (RecalL0025) sample are shown as solid (dashed) lines at the top of the plot, with red lines indicating the dispersion-dominated galaxy population and the green lines indicating the rotation-dominated galaxies.}
 \label{fig:fcov_pd_kappa}
\end{figure}

\note{The relationship between $\kappa_{\rm{HI}}$ and the \HI covering fraction \note{is explored} in Fig.~\ref{fig:fcov_pd_kappa}, where the median covering fraction is skewed to significantly higher values for galaxies with high $\kappa_{\rm{HI}}$ values at $z$~$= 1$. Here we used the rotation/dispersion-dominated classification detailed in Section \ref{sec:HI_kinematics_method}. In other words, at this redshift, the DLA covering fraction is greater for \HI discs than for dispersion-dominated \HI morphologies. We find that there is good agreement between the median f$_{\rm{cov, R_{\rm{200}}}}$ values at $z$~$= 1$ in different EAGLE boxes for both kinematic populations. From Fig.~\ref{fig:fcov_pd_kappa} we can see at $z$~$= 0$, our galaxy sample is dominated by galaxies with low covering fractions even in the sub-sample of $\kappa_{\rm{HI}} \ge 0.67$ galaxies. However, the latter displays a higher covering fraction tail that is not present for the dispersion-dominated galaxies. This tail is more prominent for the sample of galaxies taken from the RecalL0025 EAGLE box. Also, the median f$_{\rm{cov, R_{\rm{200}}}}$ values are systematically higher in RecalL0025 compared with the RefL0100 simulation.}

Overall, from Fig. \ref{fig:fcov_pd_kappa} we know that the higher covering fractions/ higher relative impact parameters DLAs seen in Fig. \ref{fig:fcov_prop} are likely to be associated with rotationally dominated \HI morphologies, supporting the hypothesis that the absorption detection in \citet{Allison2020} was associated with a large \HI disc. 

In the next section we will explore the properties of the DLAs themselves and how these vary with $z$.

\subsection{Investigation of DLA properties}\label{sec:DLA_props}
\begin{figure}
 \includegraphics[trim=2mm 2mm 1mm 0mm,clip,width=0.95\columnwidth]{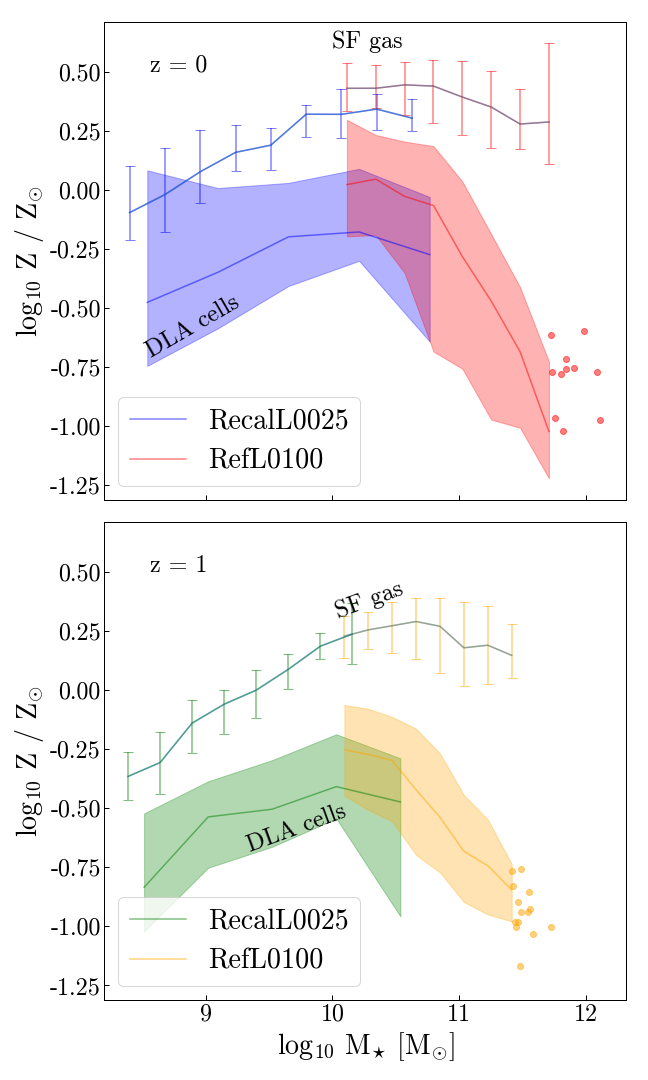}
 \caption{The mean DLA metallicity (calculated using 3D cells, see Section \ref{sec:DLA_method}) as a function of M$_{\rm{\star}}$ at $z$~$= 0$ (upper plot) and $z$~$= 1$ (lower plot) for the RefL0100 (red and orange) and RecalL0025 (blue and green) EAGLE simulations. The solid lines are the median values for both simulations, while the  16th -- 84th percentile regions are indicated by the shaded regions. We have also included the median values of the metallicity of the star-forming gas in central galaxies for the different stellar mass bins (following the same colour scheme), again with the  16th -- 84th percentile ranges shown as vertical lines.}
 \label{fig:Zvol_mstar}
\end{figure}

We plot the mean metallicities of the DLA grid cells used in the calculation of f$_{\rm{vol}}$ (see Section \ref{sec:DLA_method}), against the stellar mass of galaxies in our sample in Fig.~\ref{fig:Zvol_mstar}. We have also included the mean metallicity of star-forming gas in each galaxy as a reference. The mean metallicity of the DLA grid cells follows the same trends as the total gaseous component of the galaxies; the higher resolution simulation has systematically lower metallicities than RefL0100. Furthermore, there is an apparent disparity in the trends at low-stellar-mass ($<$ 10$^{10}$ M$_{\rm{\odot}}$) and high-stellar-mass ($>$ 10$^{10.5}$ M$_{\rm{\odot}}$). Regardless of redshift, galaxies with M$_{\rm{\star}} < 10^{10}$ M$_{\rm{\odot}}$ have DLA metallicities that show a gradual increase with M$_{\rm{\star}}$. Both boxes show a turnover between 10$^{10}$ -- 10$^{10.5}$ M$_{\rm{\odot}}$ \note{and} a subsequent negative correlation between Z$_{\rm{vol, R_{\rm{200}}}}$ at higher-stellar-mass. The mean metallicity of the DLA cells is uniformly lower than that of the star-forming gas, but this disparity increases with stellar mass at both redshifts.

These results hint that DLAs in high- and low- mass galaxies may trace different galaxy \report{evolutionary} processes -- for example, the relatively metal-poor DLAs in the low-stellar-mass ($< 10^9$ M$_{\odot}$) galaxies are likely to be associated with the smooth accretion of low-metallicity gas, while higher-metallicity DLAs in galaxies with M$_{\star} \sim$ 10$^{10}$ M$_{\odot}$ could be associated with \report{enrichment via stellar feedback}, the re-accretion of recycled gas or the accretion of metal-rich material from galaxy mergers. Previous works suggest the dominant accretion mechanism of \HI is smooth accretion, or diffuse accretion of the ionized IGM \citep{Bauermeister2010}, rather than the accretion of metal-rich material from stripped satellite galaxies. This was also found using EAGLE galaxies in \citet{Crain2017}, where the authors compare the accretion rates of \HI via smooth accretion and mergers, finding the former to dominate over our redshift range of interest (z $< 1$). \note{Additionally, \citet{Wright2020} finds that gas accretion (all gas, not just HI) comes mostly from smooth accretion of pristine gas across halo mass bins ranging from 10$^{9}$ -- 10$^{13}$ M$_{\odot}$ at $z$~$= 1.1$ \citep[at the 60$\%$ level -- see fig. 6 of][]{Wright2020}.} 
\begin{figure*}
 \includegraphics[trim=2mm 0mm 1mm 0mm,clip,width=\textwidth]{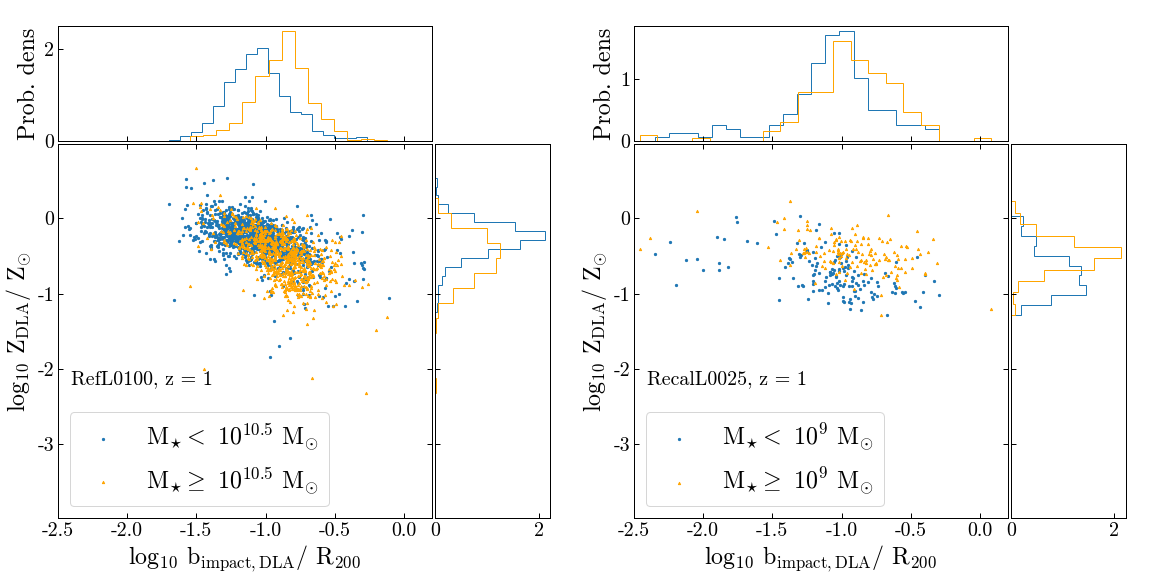}
 \caption{Main panels: the mean DLA metallicity (Z$_{\rm{DLA}}$ / Z$_{\odot}$) versus the mean impact parameter, normalised by R$_{\rm{200c}}$ (both calculated using 3D cells, see Section \ref{sec:DLA_method}) for galaxies in the RefL0100 (left column) and RecalL0025 (right column) $z$~$= 1$ samples, split into two stellar mass bins. Upper sub-panels: probability density distribution of b$_{\rm{impact}}$/R$_{\rm{200c}}$ values of the galaxies in two different stellar mass bins. Right sub-panels: probability density distribution of Z$_{\rm{DLA}}$ / Z$_{\odot}$ values of the galaxies in two different stellar mass bins.}
 \label{fig:bvol_zvol}
\end{figure*}

In order to further investigate the trends in DLA metallicity seen in Fig. $\ref{fig:Zvol_mstar}$, we now plot the mean impact parameter of the DLA cells in each galaxy (again using the 3D grid which was used to calculate f$_{\rm{vol}}$) as a function of Z$_{\rm{vol, R_{\rm{200}}}}$ in Fig.~\ref{fig:bvol_zvol} at $z$~$= 1$. Here we also show probability density distributions of the mean impact parameters and metallicities of the galaxies in our samples. The increase of normalised impact parameter with stellar mass is evident, particularly for the RefL0100 run. On the other hand, the two samples show differing trends in metallicity and stellar mass -- the RecalL0025 run shows an increase in Z$_{\rm{vol, R_{\rm{200}}}}$ between the higher and lower-stellar-mass bins ($<$ 10$^{9}$ M$_{\odot}$ and $\geq$ 10$^{9}$ M$_{\odot}$ respectively), while the lower-resolution, larger RefL0100 simulation shows a reduction in the peak of the metallicity distribution with stellar mass (between stellar mass bins $<$ 10$^{10.5}$ M$_{\odot}$ and $\geq$ 10$^{10.5}$ M$_{\odot}$). The same trends are \report{more pronounced} at $z$~$= 0$ (we do not include this plot for brevity). All stellar mass bins and both simulations show a reduction in DLA metallicity with impact parameter, although this trend is more significant in the intermediate resolution, larger EAGLE box (RefL0100). \report{However, we know from Fig.~\ref{fig:Z_rad} that the metallicity of the total gas in the galaxy is not converged in the simulation at the 10$\%$ level and hence it is unclear whether the differences in DLA properties between the simulations is driven by something physical or are numerical in origin.}

Despite the positive trend between relative impact parameter and M$_{\rm{\star}}$ seen at $z$~$= 1$ (Fig. \ref{fig:bvol_zvol}) \note{in the reference simulation (and to a lesser degree in RecalL0025)}, we note that the 2D covering fraction (f$_{\rm cov}$, calculated using face-on galaxies) of galaxies instead shows a slight reduction with halo mass at $z$~$= 1$ (top plot, Fig. \ref{fig:fcov_prop}). To investigate whether this is due to projection effects, we plotted the mean impact parameter as calculated using the 3D cells, b$_{\rm{impact, 3d}}$, minus the 2D impact parameter, calculated using a face-on orientation of the galaxies, as a function of stellar mass. Should this difference be large, this would imply significant projection effects. We can see the relative difference between the two values (normalised by R$_{\rm{200c}}$) is small until $\sim$ 10$^{10}$ M$_{\odot}$ at both redshifts. \note{At higher-stellar-masses}, there is a positive correlation between M$_{\rm{\star}}$ and the relative difference between impact parameters, indicating projection effects are having more of an impact on the observed properties of DLAs \note{in massive galaxies. However the extent of this impact appears to be resolution-dependent, as RefL0100 and RecalL0025 are offset from each other (RecalL0025 shows a sharper upturn)}. 
\begin{figure}
 \includegraphics[trim=2mm 3mm 1mm 0mm,clip,width=0.95\columnwidth]{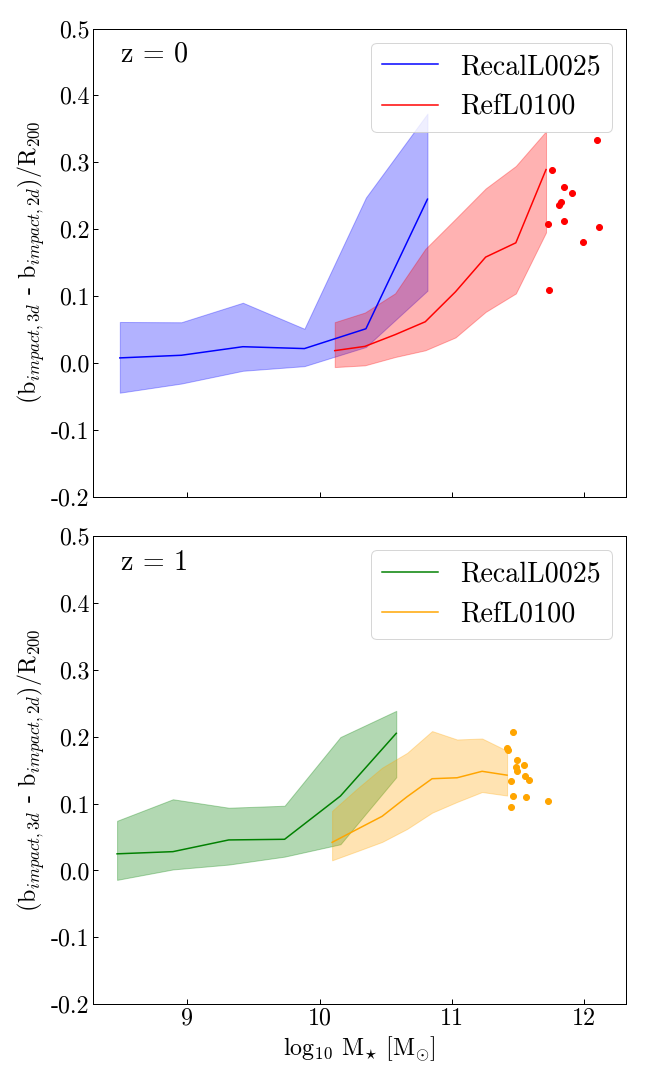}
 \caption{The difference between the impact parameters calculated using 3D cells (b$_{\rm{impact, 3d}}$) and a 2D face-on projection of the galaxy (b$_{\rm{impact, 2d}}$), normalised by R$_{\rm{200c}}$, as a function of M$_{\rm{\star}}$ at $z$~$= 0$ (upper plot) and $z$~$= 1$ (lower plot) for the RefL0100 (red, orange) and RecalL0025 (blue, green) EAGLE simulations. The solid lines are the median values for both simulations, while the shaded areas indicate the  16th -- 84th percentiles. Where there were fewer than 10 galaxies per stellar mass bin, the individual galaxies were plotted as points.}
 \label{fig:bdiff_mstar}
\end{figure}

\begin{figure}
 \includegraphics[trim=2mm 3mm 1mm 0mm,clip,width=0.95\columnwidth]{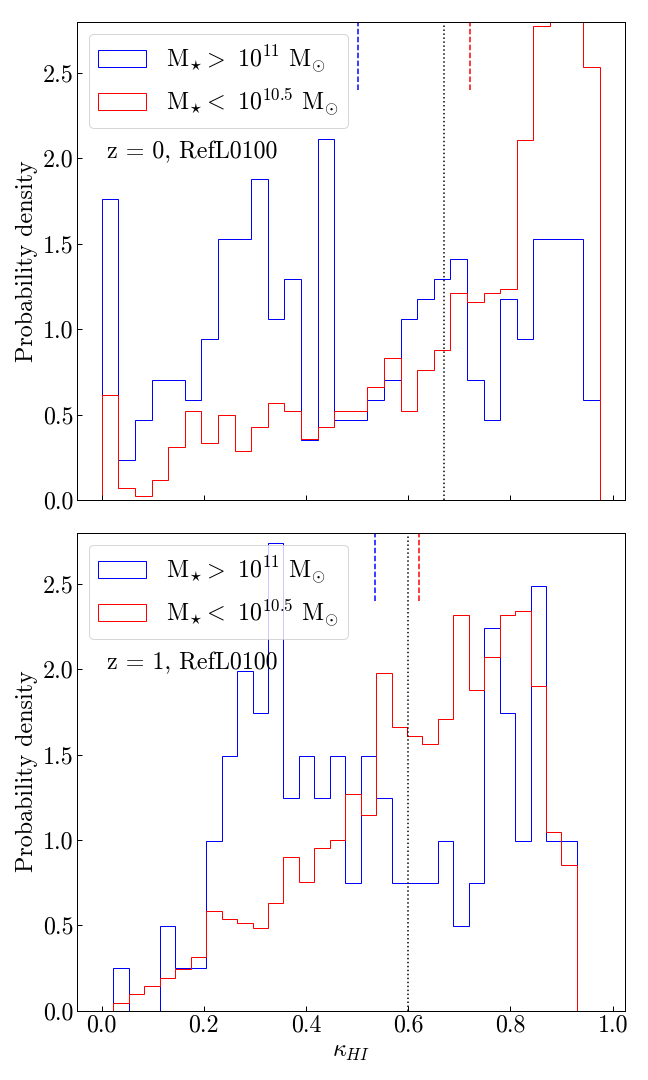}
 \caption{The normalised probability distributions of the \HI rotation parameters, $\kappa_{\rm{HI}}$, of galaxies in two stellar mass bins--$> 10^{11}$ M$_{\rm{\odot}}$ (blue) and $< 10^{10.5}$ M$_{\rm{\odot}}$  (red)--for the RefL0100 EAGLE simulations at $z$~$= 0$ (upper plot) and $z$~$= 1$ (lower plot). The vertical dotted line indicates the values of $\kappa_{\rm{HI}}$ above which we consider the \HI in a galaxy to be rotation-dominated (this is redshift dependent).}
 \label{fig:kappaHI_mstar}
\end{figure}

In order to investigate \note{how the HI} morphology, along with its kinematics, varies above and below the turnover \note{stellar} mass (10$^{10.5}$ M$_{\rm{\odot}}$ at $z$~$= 1$ and $z$~$= 0$ in the RefL0100 EAGLE simulation), we plotted the probability distribution of $\kappa_{\rm{HI}}$ (the \HI rotation parameter) for galaxies in two stellar mass bins: those with M$_{\rm{\star}} < 10^{10.5}$ M$_{\rm{\odot}}$ and those with M$_{\rm{\star}} > 10^{11}$ M$_{\rm{\odot}}$ (Fig.~\ref{fig:kappaHI_mstar}). Galaxies in the lower-stellar-mass bin have higher $\kappa_{\rm{HI}}$ values, with the mean located above 0.67 at $z$~$= 0$ and 0.6 at $z$~$= 1$, our boundary between a disc/dispersion-dominated morphology (see Section \ref{sec:HI_kinematics_method}). On the other hand, the galaxies with M$_{\rm{\star}} >$ 10$^{11}$~\note{M$_{\odot}$} show lower $\kappa_{\rm{HI}}$ values, indicative of a more irregular \HI morphology. This disparity in $\kappa_{\rm{HI}}$ is accentuated at $z$~$= 0$. Therefore, Fig.~\ref{fig:kappaHI_mstar} also supports the hypothesis that galaxies with larger stellar masses show a more irregular, dispersion-dominated \HI distribution, with DLAs at higher impact parameters relative to R$_{\rm{200c}}$.

\subsection{Relative contribution of the CGM/ISM}\label{sec:CGM}
In Section \ref{sec:HI_mass} we saw a significant mass of \HI is around the galactic disk, most likely in the CGM. It is interesting to explore whether or not this translates into a population of CGM DLAs that are likely to be picked up by absorption surveys such as FLASH. We investigate the relative contribution of the CGM to the covering fraction of DLAs, and how this varies with both galaxy properties and redshift. We followed the method described in Section \ref{sec:cgm_method} in order to split our DLA grid cells into either ISM or CGM. 

\begin{figure}
 \includegraphics[trim=6mm 5mm 22mm 0mm,clip,width=0.95\columnwidth]{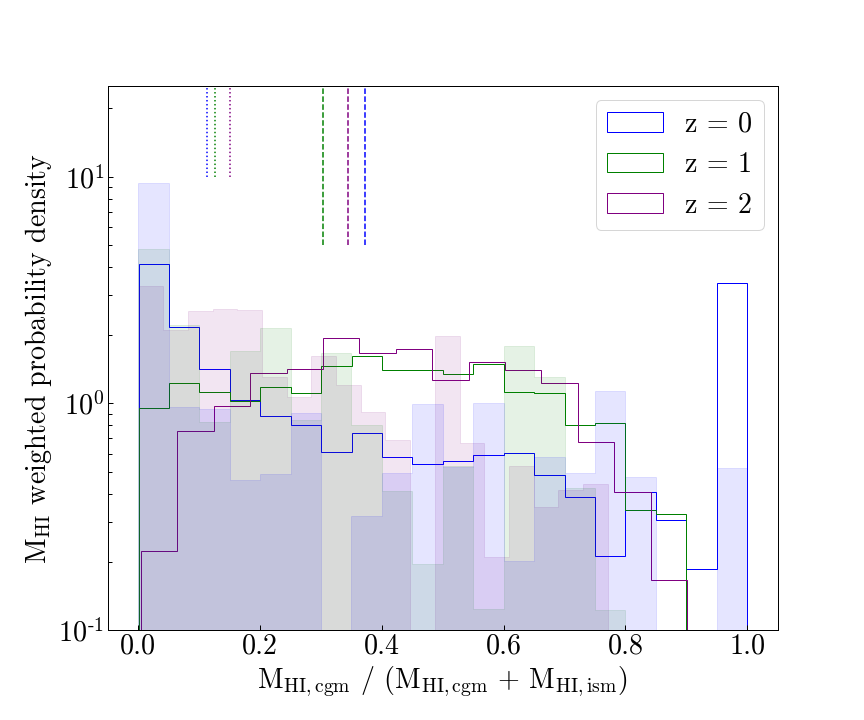}
 \caption{Probability density distribution of the ratio of the \HI mass in the CGM compared with the total \HI mass in both the ISM and the CGM (f$_{\rm{cgm, HI}}$). The distributions are shown for redshift 0, 1 and 2 in blue, green and purple respectively. The filled histograms show the results from the RecalL0025 simulation, while the others are from the RefL0100 EAGLE box. The dashed vertical lines indicate the median of each RefL0100 distribution, while the shorter, dotted vertical lines indicate the median of the galaxies sampled from the RecalL0025 simulation.}
 \label{fig:pd_CGM_HI}
\end{figure}

Initially, we focus on the global \HI budget in our sample of galaxies and how the relative contribution of the CGM and ISM varies with redshift. The upper plot of Fig.~\ref{fig:pd_CGM_HI} shows the \HI-mass-weighted histogram of the ratio of the \HI mass in the CGM compared with the total \HI mass in both the ISM and the CGM (f$_{\rm{cgm, HI}}$). We can see the \HI mass in the ISM dominates over the CGM inside galaxies in our sample at all redshift studied; the \HI-mass-weighted mean f$_{\rm{cgm, HI}}$ value is $\leq 0.37$ in both simulation boxes between$z$~$= 0 - 2$. However, the distribution of f$_{\rm{cgm, HI}}$ values varies significantly with redshift -- with the mean of the distribution increasing from 0.11 to 0.15 between $z$~$= 0 - 2$ in the RecalL0025 simulation. On the other hand, the RefL0100 shows an initial reduction in the mean between $z$~$= 0$ and $z$~$= 1$; at $z$~$= 0$ the majority of systems have a very low value for f$_{\rm{cgm, HI}}$, however there also exists a population of high \HI mass systems with high f$_{\rm{cgm, HI}}$ values that account for a significant portion of the total \HI mass budget in the galaxy sample and hence skew the mean towards higher f$_{\rm{cgm, HI}}$. On the other hand, although the median (non-\HI-mass-weighted) f$_{\rm{cgm, HI}}$ value increases between $z$~$= 0$ and $z$~$= 1$, the same high \HI mass/ high f$_{\rm{cgm, HI}}$ population of galaxies does not exist at this redshift, hence the \HI-mass-weighted mean actually decreases from 0.37 to 0.3 (before increasing to 0.35 at $z$~$= 2$). 

\begin{figure}
 \includegraphics[trim=6mm 5mm 22mm 0mm,clip,width=0.95\columnwidth]{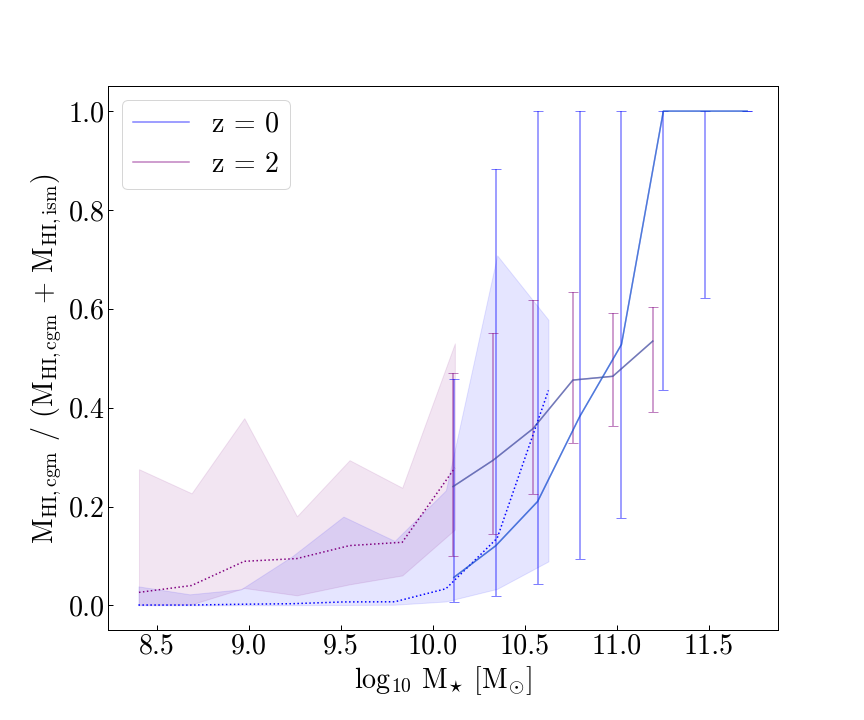}
 \caption{The median \HI mass in the CGM compared with the total \HI mass in both the ISM and the CGM (Mf$_{\rm{cgm, HI}}$) in bins of stellar mass at redshifts 0 (blue) and 2 (purple) in the RecalL0025 (dashed lines) and RefL0100 (solid lines) EAGLE simulations. The solid vertical lines/shaded regions mark the  16th -- 84th percentiles of the RefL0100/RecalL0025 galaxy samples.}
 \label{fig:CGM_HI_Mstar}
\end{figure}

There is a clear difference in the distributions of the two EAGLE boxes in Fig.~\ref{fig:pd_CGM_HI}, with RefL0025 showing lower \HI CGM mass fractions than RefL0100. While this is true at all redshifts, the disparity between the mean CGM \HI mass fraction is greatest at $z$~$= 0$. In order to ascertain whether or not this disparity arises due to the fact the two boxes are sampling galaxy populations with different halo/stellar masses, we investigate the dependence of f$_{\rm{cgm, HI}}$ on stellar mass in Fig. \ref{fig:CGM_HI_Mstar}. Here we only plot $z$~$= 0$ and $z$~$= 2$ for clarity. We can see that f$_{\rm{cgm, HI}}$ increases with increasing stellar mass in all simulation boxes at both redshifts beyond 10$^{10}$ M$_{\odot}$, however the gradient is steeper at lower redshift. At $z$~$= 0$, f$_{\rm{cgm, HI}}$ is approximately 0 in all mass bins below 10$^{10}$ M$_{\odot}$. By contrast, at $z$~$= 2$ the \HI CGM mass fraction increases gradually with stellar mass until 10$^{10}$ M$_{\odot}$, when the gradient steepens significantly. Given the change in the f$_{\rm{cgm, HI}}$ -- M$_{\star}$ relation occurs at 10$^{10}$ M$_{\odot}$ (the location of the turnover in the GSMF) these results point to AGN feedback playing \report{a role} in the \HI CGM mass fraction. This was also hinted at in Fig. \ref{fig:gas_phase}, where a significant mass of \HI was heated to higher temperatures (10$^{5}$ K) in higher mass galaxies. 

\begin{figure}
 \includegraphics[trim=0mm 28mm 17mm 40mm,clip,width=0.95\columnwidth]{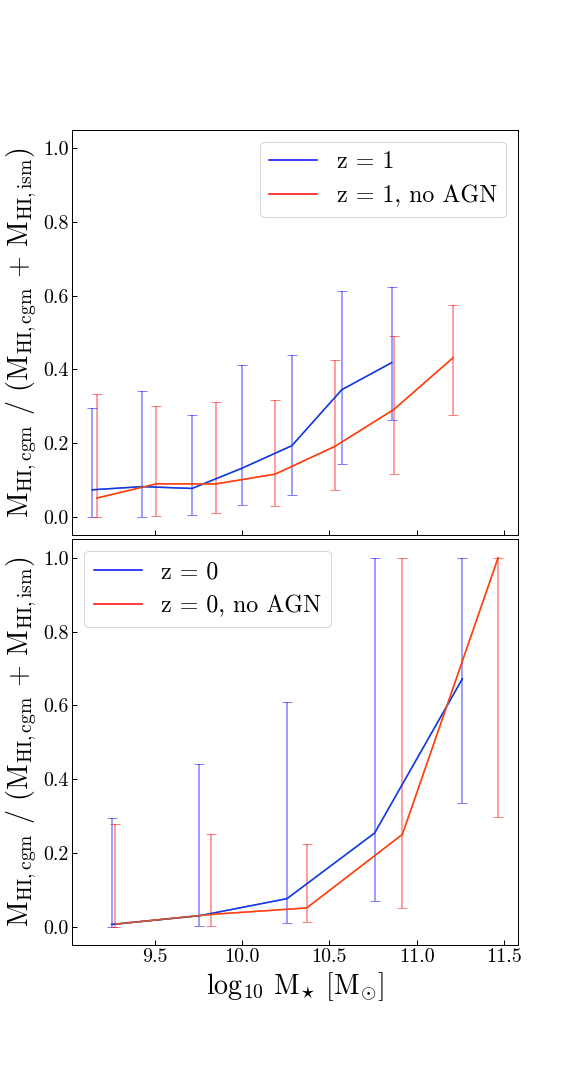}
 \caption{\report{The median \HI mass in the CGM compared with the total \HI mass (f$_{\rm{cgm, HI}}$) in stellar-mass bins at redshifts 1 (upper panel) and 0 (lower panel) in the RefL0050N0752 (blue lines) and NoAGNL0050N0752 (red lines) EAGLE simulations. The solid vertical lines mark the 16th -- 84th percentiles of the galaxy samples.}}
 \label{fig:HI_CGM_AGN}
\end{figure}

\report{In order to investigate our hypothesis that the \HI CGM mass fraction is intrinsically linked to AGN feedback, we re-plotted Fig. \ref{fig:CGM_HI_Mstar} at $z$~$= 0$ and $z$~$= 1$ for the RefL0050N0752 (blue) and NoAGNL0050N0752 (red) EAGLE simulations (Fig. \ref{fig:HI_CGM_AGN}). Focussing first at $z$~$= 1$ (upper panel), the addition of AGN feedback acts to increase the median \HI CGM mass relative to the total \HI mass (CGM and ISM) for galaxies with stellar mass greater than 10$^{9.75}$ M$_{\odot}$. Without AGN feedback, the f$_{\rm{cgm, HI}}$ -- M$_{\star}$ relation also steepens for high-stellar-mass galaxies, however the gradient is shallower and begins to increase at a stellar mass $\sim$ 0.5 dex above that for the simulation that includes AGN feedback. Focussing instead at $z$~$= 0$ (lower panel), the difference in the median f$_{\rm{cgm, HI}}$ value in different stellar mass bins with/without AGN feedback is less pronounced. The principle effect of including AGN feedback is the significant increase in scatter in the f$_{\rm{cgm, HI}}$ -- M$_{\star}$ relation towards higher f$_{\rm{cgm, HI}}$ values, seen for galaxies with stellar masses below 10$^{10.5}$ M$_{\odot}$. These results point to the importance of AGN feedback when considering the cold/cool phases of the CGM.}

The offset in f$_{\rm{cgm, HI}}$ value seen in Fig.~\ref{fig:pd_CGM_HI} between simulations of different resolution is largely removed when accounting for stellar mass in Fig.~\ref{fig:CGM_HI_Mstar}, particularly at higher redshift. At $z$~$= 0$, galaxies in the RecalL0025 EAGLE box with stellar mass $> 10^{10}$ M$_{\odot}$ have a median \HI CGM fraction that is greater than galaxies in the intermediate resolution RefL0100 EAGLE box. This could either be due to the different feedback efficiencies adopted, in particular for the AGN feedback, or the change in numerical resolution. For example, \citet{VandeVoort2019}
argue that increasing the numerical resolution of the simulation results in a higher \HI mass inside the CGM.

\begin{figure}
 \includegraphics[trim=7mm 5mm 6mm 0mm,clip,width=0.95\columnwidth]{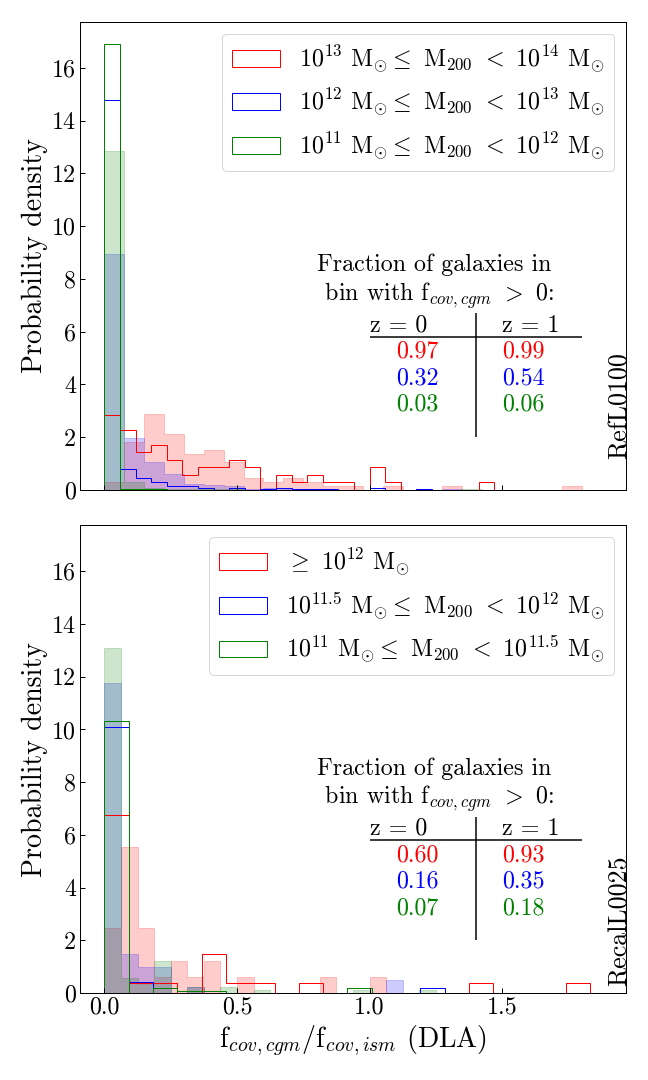}
 \caption{Normalised probability density distributions of the ratio f$_{\rm{cov, cgm}}$/ f$_{\rm{cov, ism}}$ in M$_{\rm{200}}$ bins at $z$~$= 0$ (non-filled histograms) and $z$~$= 1$ (shaded histograms) in the RefL0100 EAGLE simulation (upper plot) and the RecalL0025 (lower plot). Here we only plot galaxies with non-zero CGM covering fractions, while the fraction of the total number of galaxies in each mass range with a non-zero CGM fraction is listed on each plot.}
 \label{fig:pd_fcovcgm}
\end{figure}

Therefore, given that the CGM contains a significant portion of the \HI mass budget inside higher mass haloes, we might expect the covering fraction of CGM DLAs to be enhanced with respect to the ISM covering fraction in these galaxies. We investigate this in  Fig.~\ref{fig:pd_fcovcgm}, which shows the probability density distributions of the fraction f$_{\rm{cov, cgm}}$/ f$_{\rm{cov, ism}}$ at $z$~$= 1$ and $z$~$= 0$, using 3 different M$_{\rm{200}}$ bins (see Fig. \ref{fig:pd_fcovcgm}) in both simulations (RefL0100 and RecalL0025). We do indeed see a shift towards higher values of the CGM/ ISM covering fraction ratio with increased stellar mass. This is present at both redshifts, but is more pronounced at $z$~$= 1$. This is despite the fact we saw in Fig. \ref{fig:CGM_HI_Mstar} that the gradient in the M$_{\rm{HI, cgm}}$ -- M$_{\star}$ relation is shallower at higher redshift. In addition, the fraction of galaxies with a non-zero DLA-CGM covering fraction increases with redshift. As expected from Fig. \ref{fig:CGM_HI_Mstar}, the fraction of galaxies with non-zero CGM covering fractions is significantly reduced in the RecalL0025 simulation compared with the RefL0100 sample, primarily due to the fact the simulation boxes are sampling different mass haloes/galaxies.

Our results indicate that the CGM plays an important role when considering the properties of DLAs in galaxies with high stellar/halo mass. Furthermore, the contribution of these higher impact parameter, lower-metallicity CGM DLAs to blind \HI observation surveys such as FLASH is likely to be enhanced at higher redshift. The mass and state of the \HI in the CGM is subject to the detailed modelling of feedback physics, in particular AGN feedback. In this way, DLAs offer a way of further understanding and placing constraints on galactic feedback models.

\subsection{The Cold Neutral Medium}
\begin{figure*}
 \includegraphics[trim=0mm 4mm 0mm 0mm,clip,width=\textwidth]{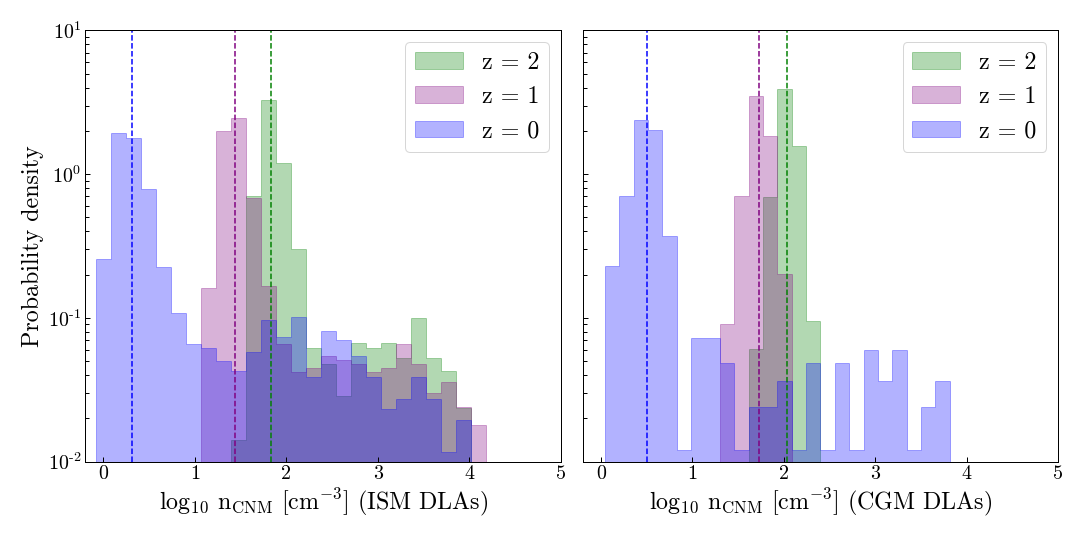}
 \caption{Normalised probability density distributions of the mean n$_{\rm{CNM}}$ calculated using our \HI model (see Section \ref{sec:HI_method}) for DLAs in the ISM (left plot) and CGM (right plot) for galaxies in the RefL0100 EAGLE simulation as a function of redshift ($z$~$= 0$ -- blue, $z$~$= 1$ -- purple, $z$~$= 0$ -- green). The medians of each distribution are indicated using the dashed lines, coloured according to redshift.}
 \label{fig:pd_ncnm}
\end{figure*}

It is important to note that since 21-cm absorption surveys are sensitive to DLAs with the highest CNM fraction, it is unclear how many of CGM DLAs presented in Section \ref{sec:CGM} would be detected in a blind \HI absorption survey such as FLASH. Moreover, our models assume either a two-phase (WNM and CNM) medium or hydrostatic equilibrium (see Section \ref{sec:HI_method} for details), neither of which may apply in the CGM. In our follow-up paper we will discuss the detailed properties and modelling of CGM DLAs, but this is beyond the scope of this paper. We investigate the redshift evolution of the CNM number densities (calculated using the outlined assumptions) for both the DLAs in the ISM and CGM in Fig. \ref{fig:pd_ncnm}.
 
Fig. \ref{fig:pd_ncnm} shows there is an increase in the peak of the n$_{\rm{CNM}}$ probability distribution with redshift, visible for both ISM and CGM DLA. This hints that the CGM DLAs are more likely to be detected in 21-cm absorption surveys at higher redshift. The ISM DLAs show a high CNM density tail that is not present for the CGM DLAs, shifting the mean n$_{\rm{CNM}}$ towards higher values. However, the median n$_{\rm{CNM}}$ for the CGM DLAs is larger than that of the ISM at all redshift. This likely arises from the fact that the ISM has a higher metallicity than the CGM, along with a greater UV flux from star-forming regions, which shifts the molecular mass fraction towards a higher value for a given CNM fraction according to our model (see Section \ref{sec:HI_method}), reducing the assumed \HI mass fraction accordingly.

\subsection{Comparing f$_{\rm{cov, R_{\rm{200}}}}$, f$_{\rm{cov, 70 kpc}}$, f$_{\rm{vol, R_{\rm{200}}}}$}
\begin{figure}
 \includegraphics[trim=6mm 5mm 22mm 0mm,clip,width=0.95\columnwidth]{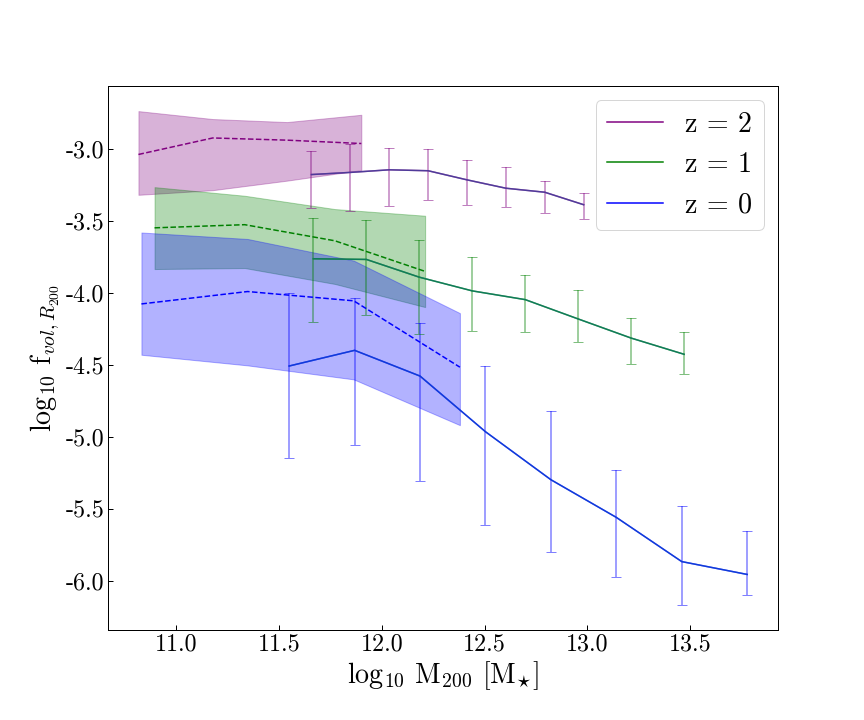}
 \caption{Figure to show the variation of f$_{\rm{vol, R_{\rm{200}}}}$ with M$_{200}$ for our sample of galaxies taken from the RefL0100 (RecalL0025) EAGLE simulation at $z$~$= 0$ (blue),$z$~$= 0.5$ (green) and $z$~$= 1$ (purple), with the median values shown as solid (dashed) lines and the  16th -- 84th percentile range shown using the capped vertical lines (shaded regions).}
 \label{fig:fvol_m200}
\end{figure}

In order to explore any differences in the trends seen for f$_{\rm{cov, R_{\rm{200}}}}$ and f$_{\rm{vol, R_{\rm{200}}}}$ (in other words between the 2D covering fraction and the volume filling fraction) with galaxy properties, we re-plotted Fig. \ref{fig:fcov_prop}, using f$_{\rm{vol, R_{\rm{200}}}}$ instead. We found no significant changes between the trends with sSFR and $\kappa_{\rm{\star}}$. Fig.~\ref{fig:fvol_m200} shows the f$_{\rm{vol, R_{\rm{200}}}}$ -- M$_{\rm{200}}$ relation. Here we see the same turnover seen in Fig.~\ref{fig:fcov_prop} with stellar mass at $z$~$= 0$. However, at $z$~$= 1$ the f$_{\rm{vol, R_{\rm{200}}}}$ declines across all halo masses, despite relatively constant f$_{\rm{cov, R_{\rm{200}}}}$ values seen at $z$~$= 1$ in Fig~\ref{fig:fcov_prop} across a wide range of halo masses. 

\begin{figure}
 \includegraphics[trim=4mm 5mm 22mm 0mm,clip,width=0.95\columnwidth]{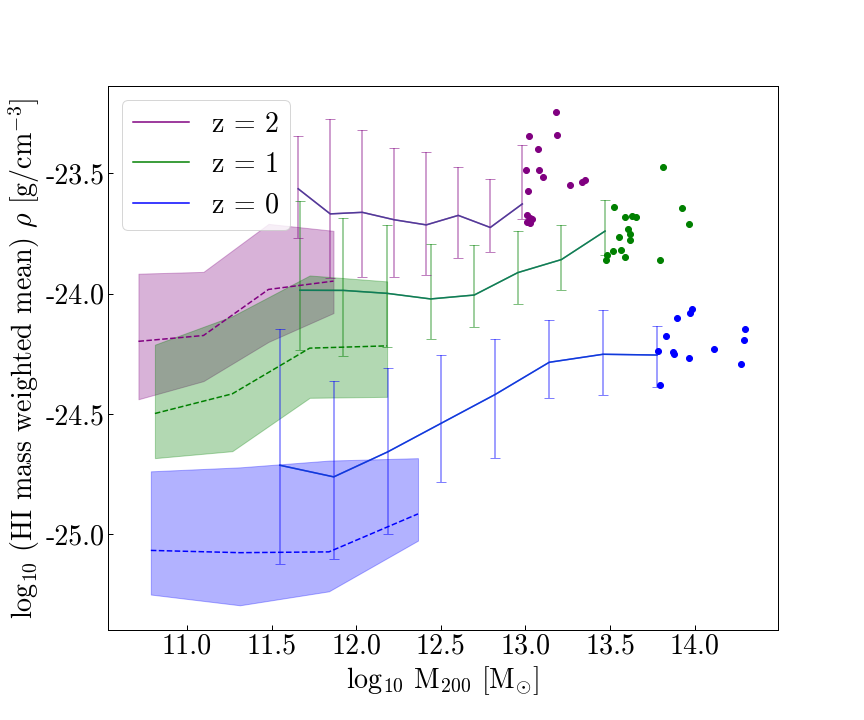}
 \caption{The mean \HI mass-weighted density of the gas inside galaxies a function of M$_{\rm{200}}$ at $z$~$= 0$ (blue), $z$~$= 1$ (green) and$z$~$= 2$ (purple). The solid/ dashed lines indicate the median values for the galaxy sample taken from the RefL0100/ RecalL0025 simulations respectively. The  16th -- 84th percentile ranges are shown using the capped vertical lines/ shaded areas for the RefL0100/ RecalL0025 simulations.}
 \label{fig:HI_mw}
\end{figure}

\begin{figure}
 \includegraphics[trim=6mm 5mm 22mm 0mm,clip,width=0.95\columnwidth]{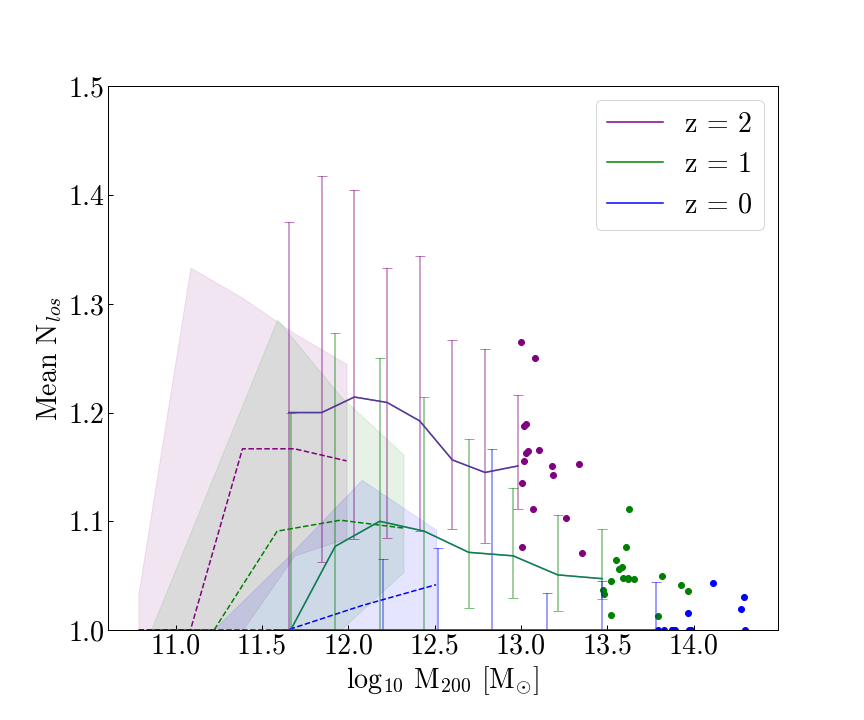}
 \caption{The number of DLA-containing cells that are summed over when calculating the column density of each pixel using the 2D face-on grid of each galaxy at $z$~$= 0$ (blue), $z$~$= 1$ (green) and$z$~$= 2$ (purple). This is a proxy for the number of DLAs per sightline. The solid/ dashed lines indicate the median values for the galaxy sample taken from the RefL0100/ RecalL0025 simulations respectively. The  16th -- 84th percentile ranges are shown using the capped vertical lines/ shaded areas for the RefL0100/ RecalL0025 simulations.}
 \label{fig:nlos}
\end{figure}

We hypothesize that the drop in f$_{\rm{vol, R_{\rm{200}}}}$ with halo mass seen at $z$~$= 1$, despite constant f$_{\rm{cov, R_{\rm{200}}}}$ values, is caused by an increase in the clumpiness of DLAs with halo mass and a corresponding drop in the number of cells containing DLAs per sightline. This would result in a drop in f$_{\rm{vol, R_{\rm{200}}}}$, despite a constant f$_{\rm{cov, R_{\rm{200}}}}$ value. We explore this hypothesis in Figures~\ref{fig:HI_mw} and \ref{fig:nlos}. In Fig.~\ref{fig:HI_mw} we show the \HI-mass-weighted mean density of the gas in each galaxy against M$_{\rm{200}}$. There is a positive correlation between $\rho$ and M$_{\rm{200}}$ at $z$~$= 0$ and $z$~$= 1$, however a plateau/ slight anti-correlation is seen at $z$~$= 2$ in the RefL0100 EAGLE simulation. This shows that \HI is on average denser in larger galaxies at z~$\leq 1$, supporting the idea that DLAs are clumpier in these higher mass systems. It is also evident from Fig. \ref{fig:HI_mw} that the mean \HI-mass-weighted density is systematically reduced in the higher resolution simulations. Since cooling is intrinsically linked to the metallicity of the gas \citep[e.g.][]{Wiersma2009}, and this is systematically lower in the higher resolution EAGLE box compared with the reference run, it follows that we would expect the mean density of the cool gas in RecalL0025 to be reduced compared with RefL0100 -- if the gas is unable to cool as efficiently, it will be more resistant to fragmentation and collapse. 

\begin{figure}
 \includegraphics[trim=6mm 5mm 22mm 0mm,clip,width=0.95\columnwidth]{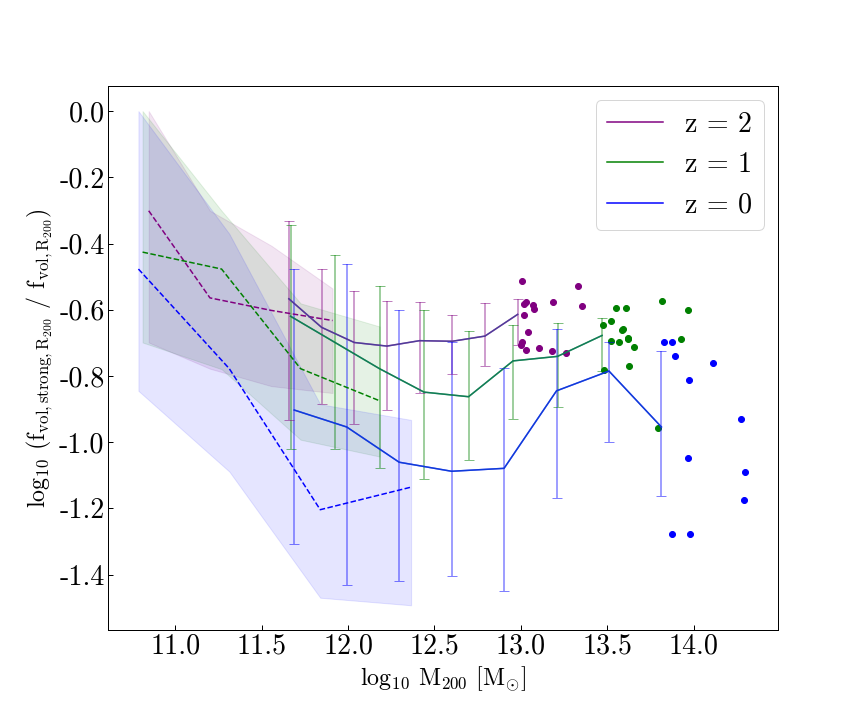}
 \caption{The ratio of the strong DLA volume-filling fraction to the total DLA volume fraction and how this changes with M$_{\rm{200}}$ at $z$~$= 1$. Here a DLA is characterised as `strong' if its \HI column density is greater than 10$^{21}$ cm$^{-2}$. The median values in halo mass bins are indicated by the solid/ dashed lines for the RefL0100/ RecalL0025 sample of galaxies. The  16th -- 84th percentile is range shown using the capped vertical lines/ shaded regions for the RefL0100 EAGLE box/ the RecalL0025 run.}
 \label{fig:z1_fvol_str}
\end{figure}

\begin{figure}
 \includegraphics[trim=4mm 0mm 30mm 5mm,clip,width=0.95\columnwidth]{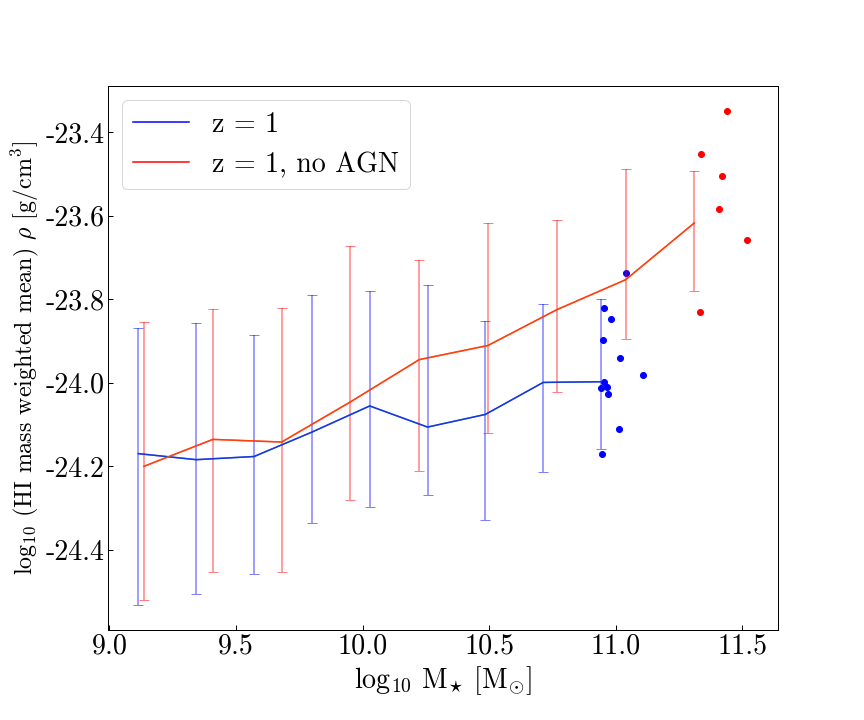}
 \caption{\report{The mean \HI-mass-weighted density of galactic gas as a function of stellar mass, using galaxies taken from the RefL0050N0752 (blue) and NoAGNL0050N0752 (red) EAGLE simulations at $z$~$= 1$. The solid lines show the median values in 10 stellar mass bins, while the vertical lines indicate the 16th -- 84th percentile range. Points are plotted for individual galaxies where the stellar mass bin contains fewer than 10 galaxies.}}
 \label{fig:z1_Hidens_agn}
\end{figure}

\begin{figure}
 \includegraphics[trim=6mm 6mm 30mm 10mm,clip,width=0.95\columnwidth]{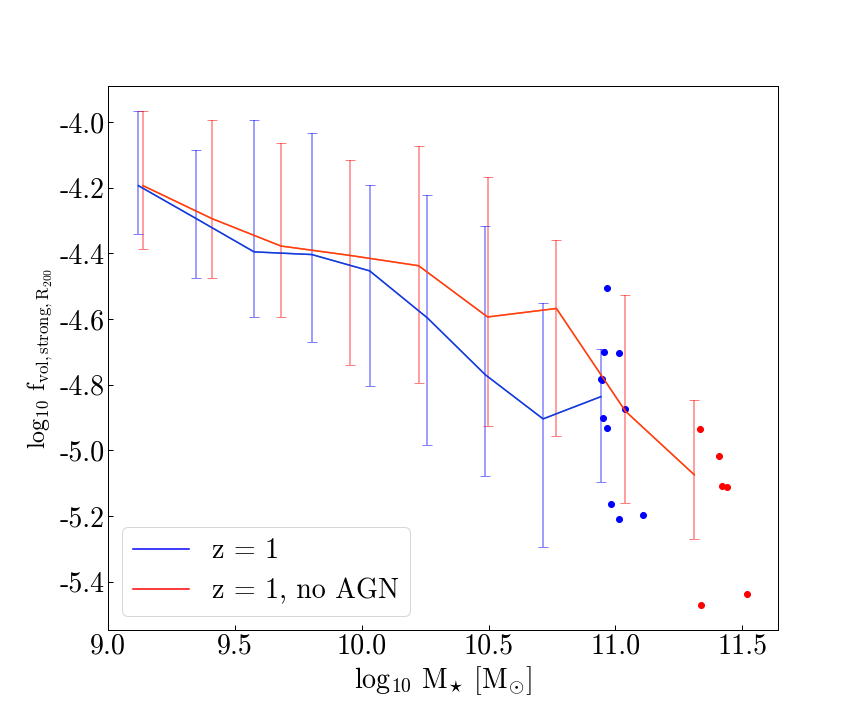}
 \caption{\report{The volume-filling fraction (measured using an aperture of R$_{\rm{200c}}$) of strong DLAs (defined as those with \HI column densities greater than 10$^{21}$ cm$^{-2}$) as a function of stellar mass, plotted using galaxies taken from the RefL0050N0752 (blue) and NoAGNL0050N0752 (red) EAGLE simulations at $z$~$= 1$. The solid lines show the median values in 10 stellar mass bins, while the vertical lines indicate the 16th -- 84th percentile range. Points are plotted for individual galaxies where the stellar mass bin contains fewer than 10 galaxies.}}
 \label{fig:z1_fvol_strong_agn}
\end{figure}

\begin{figure}
 \includegraphics[trim=6mm 5mm 22mm 0mm,clip,width=0.95\columnwidth]{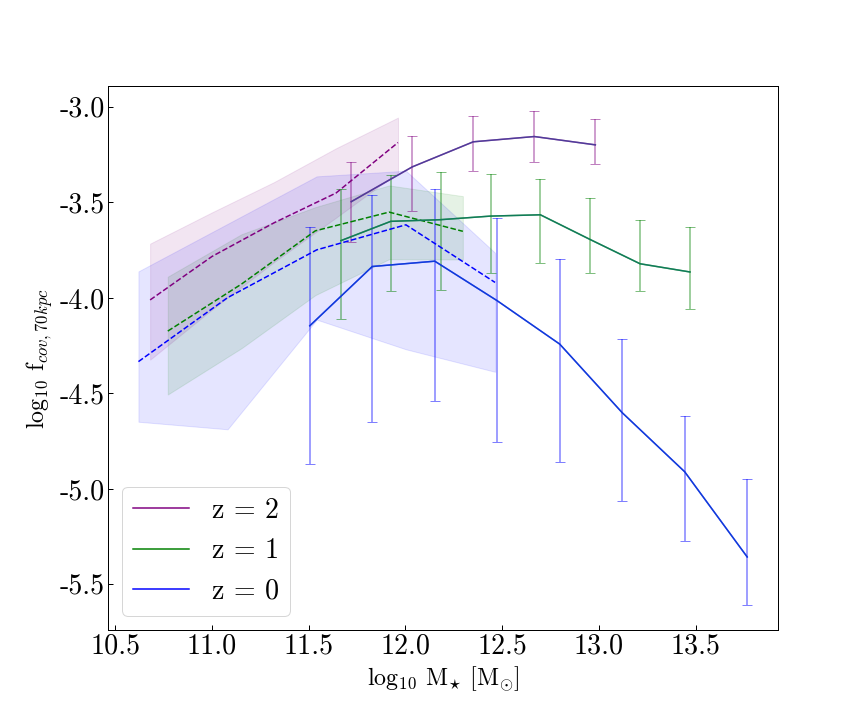}
 \caption{f$_{\rm{cov, 70 kpc}}$ as a function of M$_{200}$ for our sample of galaxies taken from the RefL0100 (RecalL0025) EAGLE simulation at $z$~$= 0$ (blue), $z$~$= 1$ (green) and $z$~$= 1$ (purple), with the median values shown as solid (dashed) lines and the  16th -- 84th percentile range shown using the capped vertical lines (shaded regions).}
 \label{fig:z1_fcov70}
\end{figure}

Fig.~\ref{fig:nlos} shows the mean number of DLA-containing cells per sightline (defined as the 2D grid cells used when calculating f$_{\rm{cov}}$; referred to as N$_{{los}}$ hereafter) against M$_{\rm{200}}$. We also see a large scatter at halo masses $\lesssim 10^{12.25}\,\rm M_{\odot}$ (particularly at high redshift). However beyond $\sim$ 10$^{12.25}$ M$_{\rm{\odot}}$ there is a negative trend between the two quantities at $z$~$= 1$ -- showing that the higher mass galaxies, on average, contain fewer DLA cells per individual sightline, which would result in a lower f$_{\rm{vol}}$ fraction, despite having a similar f$_{\rm{cov}}$ value to their lower mass counterparts. This trend extends to halo masses of 10$^{12}$ M$_{\odot}$ at $z$~$= 2$ and is also steeper at this redshift. The median N$_{los}$ for galaxies at $z$~$= 0$ is 1 across all halo masses in the RefL0100 EAGLE simulation, however, the higher resolution box shows an increase with halo mass at around 10$^{12}$ M$_{\odot}$. These results support our hypothesis that at $z$~$= 1$, the DLAs in both low and high mass galaxies inside our sample are clumpier, or less spatially extended, than those in haloes of intermediate halo masses ($\sim$ 10$^{12.5}$ M$_{\odot}$).
\newline \indent We also investigate the ratio between the volume-filling fraction of strong DLAs (those with N$_{\rm{HI}}$ $>$ 10$^{21}$ cm$^{-2}$) and all DLAs and how this varies with M$_{\rm{200}}$ in Fig.~\ref{fig:z1_fvol_str}. This shows the opposite trends of Fig. \ref{fig:nlos}, with an initial negative correlation between the two quantities until the turning point at M$_{\rm{200}}$~$\sim$~10$^{12}$ M$_{\rm{\odot}}$--10$^{12.5}$ M$_{\rm{\odot}}$ (depending on redshift), when the ratio starts to increase with halo mass. This indicates the relative contribution of strong DLAs to the f$_{\rm{vol, R_{\rm{200}}}}$ values increases with M$_{\rm{200}}$, which is in agreement with the idea that the DLAs are denser and clumpier in these systems, reducing f$_{\rm{vol, R_{\rm{200}}}}$, despite a corresponding increase in the global \HI mass (see Fig. \ref{fig:MHI_Mstar}) and a constant value of f$_{\rm{cov, R_{\rm{200}}}}$. 
\newline \indent \report{In Fig. \ref{fig:z1_Hidens_agn} we investigate whether or not the positive trend between halo mass and the \HI-mass-weighted density, seen in Fig. \ref{fig:HI_mw}, is impacted by AGN feedback. Here we plot the mean \HI-mass-weighted density of the gas in galaxies at $z$~$= 1$, taken from the RefL0050N0752 (blue) and NoAGNL0050N0752 (red) EAGLE simulations respectively. The addition of AGN feedback lowers the mean \HI-mass-weighted density in galaxies with stellar-masses above 10$^{10}$ M$_{\odot}$. This result is complimentary to Fig. \ref{fig:HI_CGM_AGN}; the fact the CGM \HI mass fraction of high-stellar-mass galaxies increases when AGN feedback is included implies a corresponding reduction in the mean \HI-mass-weighted density of the gas in these galaxies. Moreover, we also see the drop in the mean \HI-mass-weighted density is accompanied by a reduction in the volume-filling fraction of strong DLAs when AGN feedback is included (see Fig. \ref{fig:z1_fvol_strong_agn}). These results support our conclusion that AGN feedback impacts the physical properties of DLAs in high-mass galaxies. They are also consistent with work by \citet{Wright2021}, who use EAGLE simulations to show the addition of AGN feedback produces galaxies with lower baryon fractions and baryonic accretion rates. This helps to explain why the volume-filling fraction of strong DLAs increases in the case when AGN feedback is not included -- there is simply more cold, dense gas present in galaxies. Consequently, the AGN feedback model employed by cosmological simulations is likely an important factor when comparing the DLAs in simulated galaxies to those observed by surveys such as FLASH.}
\newline \indent We also re-plotted Fig.~\ref{fig:fcov_prop} using f$_{\rm{cov, 70 kpc}}$ instead (see Fig.~\ref{fig:z1_fcov70}). The trends in Fig.~\ref{fig:z1_fcov70} are broadly consistent with those in Fig.~\ref{fig:fcov_prop}; a turnover is still seen at $z$~$= 0$ at M$_{\rm{200}} \sim$ 10$^{12}$ M$_{\rm{\odot}}$, while at higher resolution there is a plateau in the covering fractions beyond 10$^{12}$ M$_{\rm{\odot}}$. There is also a marginally steeper gradient in the f$_{\rm{cov, 70 kpc}}$--M$_{200}$ at halo masses below 10$^{12}$ M$_{\odot}$, compared with the gradient seen using the virial radius as the aperture in Fig.~\ref{fig:fcov_prop}.  This is driven by the use of a fixed aperture across a range of halo masses, while the \HI disc extent increases with halo mass. However, the presence of the peak in the $z$~$= 0$  f$_{\rm{cov, 70 kpc}}$--M$_{200}$ relation is likely due to a significant number of the DLAs in the higher mass systems being at radii larger than 70 kpc in the CGM of the halo, in agreement with the significant total \HI mass outside 70 kpc seen in Fig.~\ref{fig:MHI_Mstar}. This result is linked to the increase in impact parameter relative to R$_{\rm{200c}}$ at higher halo masses at $z$~$= 0$ seen previously in Fig. \ref{fig:bvol_zvol}. 
\label{sec:fcov_vs_fvol}

\section{Discussion and Conclusions}\label{sec:discuss_and_concl}
This paper set out to investigate the distribution and properties of DLAs in the EAGLE cosmological hydrodynamical simulations, focussing on the redshift range 0 to 2 in order to connect with the results from the next generation of 21-cm absorption surveys. The results of this paper can be summarised as follows:
\begin{itemize}
    \item The \HI-to-stellar mass ratio increases with redshift in EAGLE galaxies, with the most significant increase seen at M$_{\rm{\star}}$ $>$ 10$^{10}$ M$_{\rm{\odot}}$ (see Fig.~\ref{fig:MHI_Mstar}). The \HI masses obtained for the lower resolution, larger EAGLE box are systematically lower than those obtained at the same stellar mass in the high-resolution box. This is due to the higher gas metallicities of the former compared to the latter at fixed stellar mass, which leads to more efficient H$_{\rm{2}}$ formation. This effect is lessened at higher redshift due to the central parts of galaxies in the high- and low-resolution boxes having similar metallicities, despite the average metallicity of gas in the galaxies being significantly higher in the lower resolution EAGLE simulation (see Fig.~\ref{fig:Z_rad}). The exact shape and normalisation of the M$_{\rm{HI}}$ -- M$_{\rm{\star}}$ relation is impacted by the chosen aperture in which the \HI content in galaxies of stellar masses $>10^{10}\rm \, M_{\odot}$ is measured (see Fig. \ref{fig:MHI_Mstar}), indicating that a significant fraction of the \HI mass is at larger radii in these galaxies. This effect is seen both at $z$~$= 0$ and $z$~$= 1$.
    \item The median DLA covering fraction obtained using an aperture of R$_{\rm{200c}}$, f$_{\rm{cov, DLA}}$, shows a non-linear dependence on halo mass at $z$~$= 0$, with a maximum at $\sim$ 10$^{12}$ M$_{\rm{\odot}}$. This maximum corresponds to the break of the GSMF both observationally and theoretically and is typically attributed to AGN\report{/stellar} feedback becoming efficient\report{/inefficient} at higher-stellar-masses. On the other hand, at $z$~$= 1$, f$_{\rm{cov, R_{\rm{200}}}}$ does not vary over a wide range of halo masses. Moreover, there is a global shift towards higher DLA covering fractions (both 2D and 3D volume filling fractions) between $z$~$= 0$ and $z$~$= 1$. 
    \item \report{On further investigation, we found the addition of AGN feedback had the largest impact on the covering fraction of DLAs at $z$~$= 0$ for host galaxies in the stellar mass range $10^{10}$ M$_{\odot}<$ M$_{\star}\leq10^{11}$ M$_{\odot}$, where it acts to reduce the DLA covering fraction (Fig. \ref{fig:fcov_prop_agn}).}
    \item Looking instead at the kinematics of the \HI, we find that EAGLE galaxies that are rotation-dominated have, on average, higher covering fractions, when compared with galaxies that are dispersion-dominated, indicating that galaxies with the highest DLA covering fractions also have orderly rotating \HI discs (Fig. \ref{fig:fcov_pd_kappa}). High-stellar-mass galaxies were also shown to have lower $\kappa_{\rm{HI}}$ values. In other words, the kinematics of the \HI is more dispersion-dominated in these galaxies. However, the difference between galaxies in different mass bins lessens with increasing redshift.
    \item In Section \ref{sec:DLA_props} we saw that DLAs in EAGLE galaxies with high-stellar-masses ($>$ 10$^{10.5}$ M$_{\rm{\odot}}$) had the highest impact parameters relative to their virial radius (Fig.~\ref{fig:bvol_zvol}), along with lower metallicities on average than those found in galaxies with stellar masses $< 10^{10.5}$ M$_{\odot}$. Furthermore, \report{these trends were not visible} in the higher-resolution RecalL0025 box; the DLAs in lower-mass galaxies ($<$ 10$^{9}$ M$_{\odot}$) \report{instead had marginally higher} metallicities than those in higher-stellar-mass galaxies ($\geq$ 10$^{9}$ M$_{\odot}$) on average. Overall, our results indicate that the DLAs in systems of differing stellar mass \report{trace different galaxy evolutionary processes}. However, it is unclear if the process driving the differences seen between the two EAGLE boxes is physical, or numerical in origin.
    \item The ISM dominates the total \HI mass budget compared to the CGM in the majority of galaxies in our samples (see Fig. \ref{fig:pd_CGM_HI}). However, we do see a trend of increasing \HI mass fraction in the CGM relative to the ISM with increasing redshift. Moreover, the CGM \HI mass fraction increases sharply with stellar mass above $\sim$ 10$^{10}$ M$_{\odot}$ at all redshifts studied here, although the trend is steeper at $z$~$= 0$ (see Fig. \ref{fig:CGM_HI_Mstar}). This trend is present in both the intermediate- and high-resolution EAGLE boxes. Since this sharp increase of CGM \HI mass fraction again occurs at the peak in the GSMF, it is likely DLAs in the CGM offer a way of further understanding and placing constraints on galactic feedback models, in particular AGN \citep[see also][for similar conclusions using semi-analytic models of galaxy formation]{Chauhan2020}. \report{Moreover, using simulations with/without AGN feedback included, we found that the presence of an AGN acts to increase the scatter in CGM \HI mass fraction for galaxies with M$_{\star} <$ 10$^{10.5}$ M$_{\odot}$ at $z$~$= 0$, while also increasing the median \HI CGM mass fraction for galaxies with M$_{\star}$ $\gtrsim$ 10$^{10}$ M$_{\odot}$ at $z$~$= 1$ (Fig. \ref{fig:HI_CGM_AGN})}. The increase in CGM \HI mass fraction in higher-stellar-mass galaxies manifests as a correlation between the ratio of the CGM-to-ISM covering fraction and M$_{\rm{200}}$ (see Fig. \ref{fig:pd_fcovcgm}). Additionally, this ratio increases with redshift.
    \item In Section \ref{sec:fcov_vs_fvol} we showed that the \HI-mass-weighted gas density increases with increasing halo mass (see Fig. \ref{fig:HI_mw}). This is accompanied with a larger relative covering fraction of strong DLAs and fewer DLAs per sightline (Figs \ref{fig:z1_fvol_str} and \ref{fig:nlos} respectively). This resulted in the f$_{\rm{vol, R_{\rm{200}}}}$ anti-correlating with halo mass at $z$~$= 1$, despite a constant f$_{\rm{cov, R_{\rm{200}}}}$ across a range of halo masses (see Figs \ref{fig:fcov_prop} and \ref{fig:fvol_m200}). A peak in the number of DLAs per sightline and a corresponding dip in the fraction of strong DLAs is seen at a halo mass between 10$^{12}$ M$_{\odot}$ -- 10$^{12.5}$ M$_{\odot}$, depending on redshift. \report{We investigated whether or not this could be impacted by the action of AGN feedback, finding that the addition of an AGN feedback model acts to reduce the mean \HI-mass-weighted density in galaxies with stellar-masses above 10$^{10}$ M$_{\odot}$ at $z$~$= 1$ (Fig. \ref{fig:z1_Hidens_agn}). There is also a corresponding reduction in the volume-filling fraction of strong DLAs in these high-stellar-mass galaxies}.
    \item In the appendices we show that the effects of altering the \HI -- H$_2$ model are not significant compared with altering the resolution of the simulation and, in doing so, the distribution of metals in galaxies along with the feedback (AGN and stellar) models. It is therefore imperative to understand and constrain the effects of both AGN and stellar feedback, along with radial metallicity distribution, on the \HI content of the CGM and ISM, in order to fully understand the results of the cosmological simulations with regard to DLAs. 
   
\end{itemize}\label{sec:discussion}
\subsection{Caveats}
One of the key results of this paper is that the high-stellar-mass galaxies in our sample have clumpier, stronger DLAs, which have a higher mean relative impact parameter and lower metallicities. As discussed in Section \ref{sec:HI_method}, we know from \citet{Rahmati2013b} that the effects of LSR on strong DLAs is highly uncertain, hence this introduces a degree of uncertainty to these results. In future work we would like to investigate the impact of LSR on these strong, clumpy DLAs. However, given the fraction of the DLAs in the CGM is higher in these galaxies, it is likely LSR will have a limited impact on the low metallicity, high impact parameter DLAs. 

Furthermore, we have consistently seen that the metallicities of the DLAs in the higher resolution boxes are systematically lower than those in the lower-resolution box, which in turn impacts the \HI/H$_{\rm{2}}$ fractions, along with cooling properties of the gas. We saw in Fig. \ref{fig:Z_rad} that the situation is actually better when we consider the radial metallicity dependence at $z$~$= 1$. Yet, how this alteration in metallicity affects our results is not easy to decipher, since rapid cooling supports both the formation of higher density \HI clumps \emph{and} the removal of \HI via the formation of H$_{\rm{2}}$. 

Both the \HI/H$_2$ breakdown and the simulations themselves are based on a spatial resolution of the order of 1 kpc \citep[the softening length for the RefL0100 run is 0.7 pkpc--see Table 2 of][]{Schaye2015}, which is at least a factor of 10 greater than the average (giant) molecular cloud/ molecular cloud width. Our results therefore average across a wider area, meaning we are missing the detailed kinematics/ mixing of the two neutral gas phases. Therefore, our DLA properties should be seen as global averages across the whole galaxy.
\subsection{Our results in context}
\report{We have shown that AGN feedback is linked to the volume-filling fraction of strong DLAs, along with the mean density of \HI gas in high-stellar-mass galaxies. In this way, our results suggest that DLAs and their physical properties offer a potential tracer for AGN feedback and its impact on the host galaxy. This links to both ongoing work into the cold component of AGN outflows \citep[e.g.][]{Lehnert2011, Morganti2016}, along with the complex interplay between AGN outflows and in-flowing gas \citep[e.g.][]{Riffel2015}. Our future work will focus on this and in particular how both AGN and stellar feedback impact not only the properties of DLAs, but also the fraction of these in the CGM.}

This paper has focused on galaxies with stellar masses $> 10^{8}$ M$_{\rm{\odot}}$ due to resolution constraints. However, we know from plots such as  Fig.~\ref{fig:z1_fvol_str} and Fig. \ref{fig:bvol_zvol} that the DLAs associated with lower-stellar-mass galaxies (those located to the left of the GSMF) and higher-mass galaxies (those beyond the peak in the GSMF) likely differ in physical origins and follow opposing trends. Therefore, pushing towards the lower end of the GSMF would offer key insights into this transition seen in the DLA population. \report{We know AGN feedback is intrinsically linked to the properties of DLAs in higher-mass systems, hence it is likely that stellar feedback also greatly impacts the \HI budget in smaller, dwarf galaxies.} In order to delve more deeply into the relative impact of different types of stellar feedback on dwarf galaxies, and in particular how this changes the DLA abundance in the CGM, cosmological zoom-in simulations are needed \citep[recent work on this includes ][]{Rhodin2018}.

The role of the CGM in galaxy evolution has been an area of intense interest in recent years, with multiple papers linking an enhancement in numerical resolution of simulations to a boost in the column density of low ions such as \HI inside the CGM \citep{VandeVoort2019,Hummels2019}. Given our results point to a significant fraction of the DLAs in high-mass galaxies originating in the CGM, it would be of interest to study these systems using cosmological zoom-in simulations. 

Recent results from the FLASH survey \citep{Allison2020} indicate the presence of large quantities of \HI at a high impact parameter ($\sim$ 17 kpc) in an intervening absorbing galaxy. Our results support the hypothesis that at high redshift, higher relative impact parameters are more likely for DLAs in galaxies of high-stellar-mass. Our results also indicate that the properties of detected DLAs are intrinsically linked with the host galaxy properties, such as mass and star formation rate. For example, DLAs in intermediate-mass galaxies (M$_{\star} \sim$ 10$^{10}$ M$_{\odot}$) are more likely to trace recycled \HI than those in lower- and higher-stellar-mass galaxies (see Fig. \ref{fig:Zvol_mstar}). 

\section*{Acknowledgements}
This research was supported by the Australian Research Council Centre of Excellence for All	 Sky Astrophysics in 3 Dimensions (ASTRO 3D), through project number	CE170100013. JRA acknowledges support from a Christ Church Career Development Fellowship. ARHS acknowledges the receipt of the Jim Buckee Fellowship at ICRAR-UWA. This work was supported by resources provided by the Pawsey Supercomputing Centre with funding from the Australian Government and the Government of Western Australia. LGS thanks Benedikt Diemer for use of his UV propagation code, described in Section \ref{sec:HI_method} of this paper. CL thanks the MERAC Foundation for a Postdoctoral Research Award. 
The Cosmic Dawn Center of Excellence is funded by the Danish National Research Foundation under the grant No. 140. 
We have used \textsc{PYTHON} for our data analysis and acknowledge the use of \textsc{MATPLOTLIB} \citep{Hunter2007} to generate the plots in this paper.

\section*{Data Availability}
This paper uses the public data release of the EAGLE simulation suite \citep{Crain2015, Schaye2015, McAlpine2016, Eagle2017} and the data underlying this article are available at \url{http://icc.dur.ac.uk/Eagle/database.php}. We also use the data from the extended GALEX Arecibo SDSS Survey \citep[xGASS;][]{Catinella2018} available at \url{https://xgass.icrar.org/data.html}, along with the publicly available photoionization tables given by \cite{Haardt2012}, available at: \url{http://www.ucolick.org/~pmadau/CUBA/HOME.html}. 




\bibliographystyle{mnras}
\bibliography{Flash_Eagle} 

\appendix
 
\section{Different \HI-H$_{\rm{2}}$ decomposition method(s)}\label{sec:app_gned}
\subsection{Gnedin 2011 breakdown}
Here we compare the results of the \citet{Krumholz2013} method (or K13) with an alternative method for calculating the ratio of molecular to neutral Hydrogen ($\rm{f_{H_2}}$), which uses the fitting formulae described in \citet{Gnedin2011} to relate the dust-to-gas ratio and the interstellar far-UV (FUV) flux to the atomic-to-molecular transition in the ISM of the simulation. \citet{Gnedin2011} derived these fitting formulae using simulations of a snapshot of a cosmological simulation taken at $z$~$= 4$ and evolved using 3D radiative transfer and a H/He chemical network, along with a H$_{2}$ formation model. We refer to this method as G11 in the text.

Fundamentally, \citet{Gnedin2011} parameterise f$_{\rm{H_2}}$ (this has the same meaning as above) as
\begin{equation}\label{Gnedin2011}
    f_{H_2} \approx \frac{1}{1 + \exp{(-4x - 3x^3)}} ,
\end{equation}
where $x$ can be expressed as 
\begin{equation}
    x = \Lambda^{3/7}\ln{(\frac{Z n_{\rm{H}}}{Z_{\odot} \Lambda n_\star})}   , 
\end{equation}
with $n_\star =$ $25$cm$^{-3}$, while
\begin{equation}
    \Lambda = \ln{(1 + g(Z/Z_\odot)^{3/7}(G_0'/15)^{4/7})}.
\end{equation}
Furthermore, the $g$ factor is given by
\begin{equation}
    g = \frac{1 + \alpha{s} + s^2}{1 + s},
\end{equation}
with $s$, $\alpha$ and $D_\star$ defined by
\begin{equation}
    s = \frac{0.04}{D_\star + (Z/Z_{\odot})},
\end{equation}
\begin{equation}
    \alpha = \frac{5G_0'/2}{1 + (G_0'/2)^2},
\end{equation}
and
\begin{equation}
     D_\star = 1.5\times10^{-3} \times \ln(1 + (3G_0')^{1.7}),
\end{equation}
respectively. Here we have substituted our value of the UV field for $G_0'$, calculated using different methods for the star-forming and non-star-forming particles, as detailed in Section \ref{sec:HI_method}.

\begin{figure}
 \includegraphics[width=0.95\columnwidth]{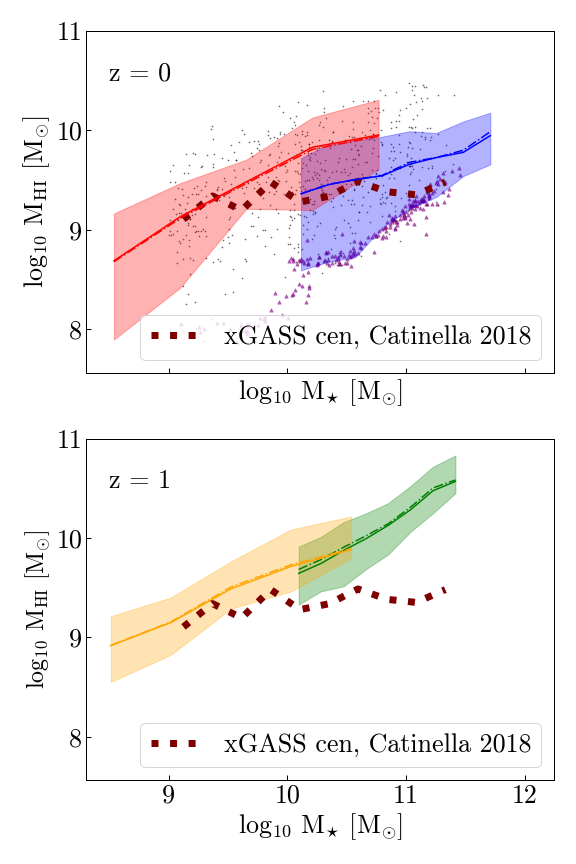}
 \caption{Plot of M$_{\rm{HI}}$--M$_{\rm{\star}}$ at $z$~$= 0$ (top plot) and $z$~$= 1$ (lower plot), as calculated for both EAGLE box RecalL0025 (red/ orange respectively) and RefL0100 (blue and green respectively) using both the G11 \HI -- H$_{2}$ breakdown model (dot-dashed lines) and the K13 model (solid lines). The solid/ dot-dashed lines indicate the median values in 10 stellar mass bins for each galaxy sample, while the shaded regions are the 16th -- 84th percentile range inside each stellar mass bin for the G11 model. Also plotted is a representative sample of the M$_{\rm{HI}}$ and M$_{\rm{\star}}$ values obtained by the xGASS survey \citep{Catinella2018} (black dots/purple triangles for detections/ non-detections) for central galaxies in the local universe, along with the median values (dashed line).}
 \label{fig:Gned_MHI_Mstar}
\end{figure}

Fig. \ref{fig:Gned_MHI_Mstar} shows there is minimal impact on the overall M$_{\rm{HI}}$--M$_{\star}$ relation by using the \citet{Gnedin2011} \HI -- H$_{2}$ breakdown method described above (from now on described as G11) instead of the theoretically motivated \citet{Krumholz2013} method (K13) described in Section \ref{sec:HI_method}. This is true for both the RecalL0025 and RefL0100, and is also seen at low and high redshift. These results show that changing the \HI model has minimal impact on the global \HI properties of resolved galaxies in the simulations.

\begin{figure}
 \includegraphics[width=0.95\columnwidth]{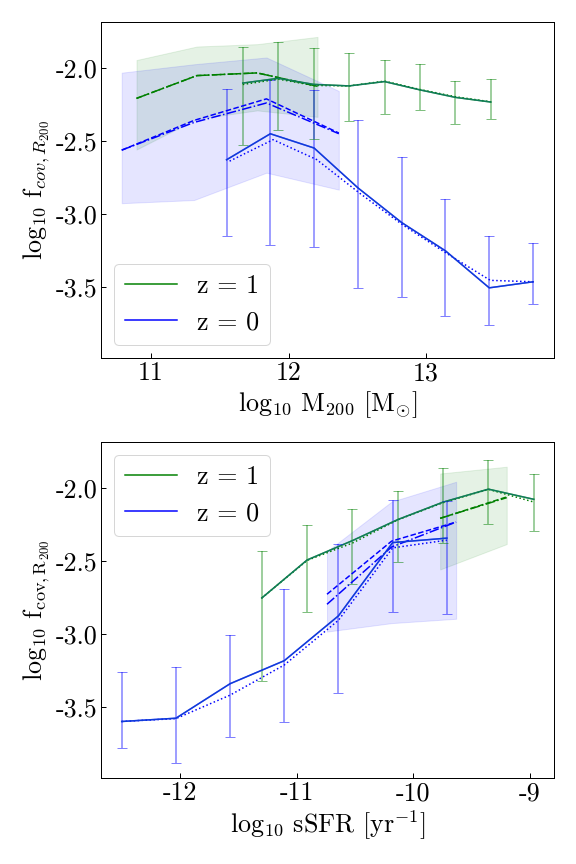}
 \caption{Plots of f$_{\rm{cov, R_{200}}}$ versus M$_{200}$ (left) and sSFR (right). In each plot, the results from $z$~$= 1$ are shown in green and those from $z$~$= 0$ are in blue. The solid/ dashed lines indicate the median values of f$_{\rm{cov, R_{200}}}$ in halo mass/ sSFR bins calculated using the K13 \HI -- H$_{2}$ breakdown method in the RefL0100/ RecalL0025 EAGLE boxes. Alternatively, the dotted/ dot-dashed lines indicate the median values calculated using the G11 model for the RefL0100/RecalL0025 simulations. The vertical lines and shaded regions indicate the 16th -- 84th percentile ranges calculated for the G11 results in the RefL0100 and RecalL0025 simulations respectively.}
 \label{fig:Gned_fcov_mstar}
\end{figure}

We next explore the impact of changing the \HI--H$_{2}$ breakdown method on the redshift evolution of the f$_{\rm{cov, R_{200}}}$--M$_{200}$ relation (Fig. \ref{fig:Gned_fcov_mstar}). We can see the trends that were central to our results in the paper -- namely the change in gradient seen between $z$~$= 0$ and  $z$~$= 1$ along with the peak at $\sim$ 10$^{12}$ M$_{\odot}$ at $z$~$= 0$ -- are still present when using the G11 method, indicating the convergence in total \HI mass/ global \HI properties seen in Fig. \ref{fig:Gned_MHI_Mstar} extends to the highest column density \HI in each galaxy.

\begin{figure}
 \includegraphics[width=0.95\columnwidth]{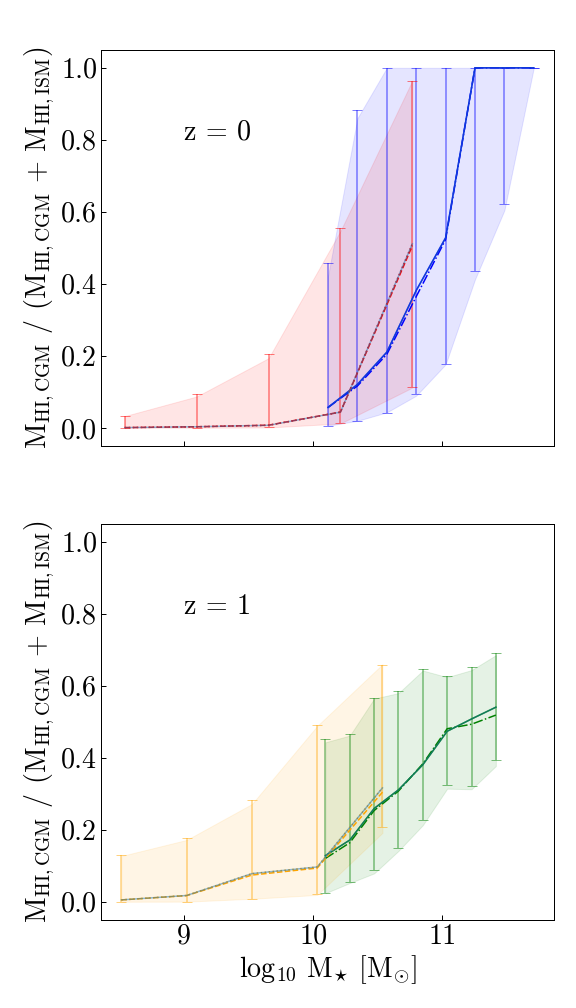}
 \caption{Plots of the mass of \HI in CGM as a fraction of the total \HI mass in the ISM and CGM, as a function M$_{\star}$ at $z$~$= 0$ (left) and $z$~$= 1$ (right). The median values across bins in stellar mass are shown as solid (dashed) lines for the results calculated using the K13 (G11) \HI--H$_2$ breakdown method. The shaded regions indicate the 16th -- 84th percentile ranges for the G11 results, while the vertical lines indicate the same for the K13 results. The results from the RecalL0025  galaxy sample are shown in red/ orange (redshift 0/1), while those taken from RefL0100 are shown in blue/ green.}
 \label{fig:Gned_cgm}
\end{figure}

Finally, we investigate the impact of the \HI -- H$_{\rm{2}}$ method on the relative mass of \HI in the CGM compared with the ISM in Fig. \ref{fig:Gned_cgm}. Here we plot the mass of \HI in the CGM as a fraction of the total \HI mass in both the CGM and ISM, as a function of stellar mass. Comparing the results from G11 to those found using the K13 method, we see they are more or less identical -- the G11 results show the same reduction in the slope of this relation with redshift, along with a global increase in the CGM \HI mass fraction seen in galaxies with stellar mass $< 10^{10}$ M$_{\rm{\odot}}$ with increasing redshift. This agreement in the CGM/ ISM \HI breakdown between different methods gives validity to our conclusions that a significant fraction of \HI lies in the CGM in high mass galaxies across the range of redshift studied here. 

\subsection{Other alterations to the \HI -- H$_2$ breakdown method}
\begin{figure}
 \includegraphics[width=0.95\columnwidth]{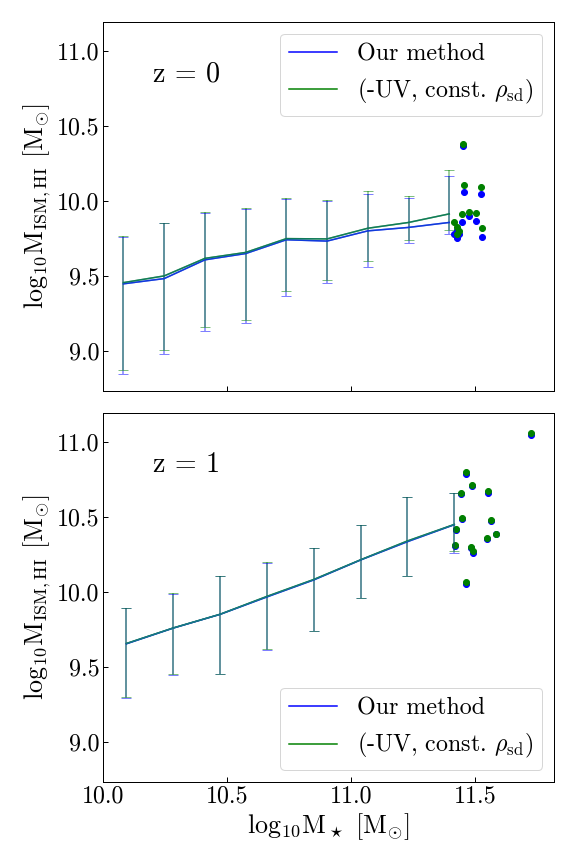}
 \caption{Plots of the median mass of \HI in ISM using bins in M$_{\star}$ at $z$~$= 0$ (top panel) and $z$~$= 1$ (lower panel) using galaxies taken from RefL0100, as calculated using two different \HI -- H$_2$ breakdown methods; the first is our fiducial method in blue, described in Section \ref{sec:HI_method}, while the second (in green) does not include the UV propagation from star forming particles and uses a constant value for the total matter density ($\rho_{\rm{sd}}$, as referenced in equation \ref{eqn:disc_pressure}). The 16th -- 84th percentile ranges are shown using the vertical, capped lines. }
 \label{fig:UV_ism}
\end{figure}

We also explored the effect of including the additional UV propagation mechanism (detailed in Section \ref{sec:HI_method}) and a non-constant background matter density ($\rho_{\rm sd}$, referenced in equation \ref{eqn:disc_pressure}), again finding our results do not change significantly, in particular the \HI masses do not vary significantly, along with the DLA covering fractions. The largest difference in the \HI masses occurs in particles we ascribe to the ISM (see Section \ref{sec:cgm_method} for our detailed ISM/ CGM breakdown); the CGM masses do not vary between the two \HI -- H$_2$ breakdown methods. This is demonstrated in Fig. \ref{fig:UV_ism}; the ISM \HI mass is reduced in high stellar mass galaxies ($>$ 10$^{11}$ M$_{\odot}$) at $z$~$= 0$. On further investigation, we found the lost \HI mass is instead considered to be H$_2$ in our fiducial method (which includes UV propagation). This is primarily because the \citet{Wolfire2003} two-phase model dominates in both the ISM and CGM (see Fig. \ref{fig:pd_ncnm}); and in this regime the CNM number density is set by the minimum pressure required to maintain the CNM alongside the WNM, which in turn is directly proportional to the UV flux (see equation \ref{eqn:nCNM_2p_simplified}). Since the additional UV flux from the propagation of UV to non-star-forming particles is greater in the ISM compared with the CGM, the associated increase in the density of the CNM is also greater, along with the assumed optical depth of the dust (see equation \ref{eqn:chi}), leading to a larger H$_2$ mass fraction and subsequently a smaller \HI mass fraction.

\bsp	
\label{lastpage}
\end{document}